\RequirePackage{fix-cm}
\documentclass[12pt]{article}
\usepackage{amsmath}
\usepackage{graphicx}
\usepackage{enumerate}
\usepackage{natbib}
\usepackage{multirow}
\usepackage{tabularx}
\usepackage{url} 
\usepackage[fontsize=11.5pt]{scrextend}
\usepackage{setspace}

\makeatletter
\newcommand*\showfontsize{\f@size{} point}
\makeatother

\newcommand{\blind}{1}

\begin{document}

\def\spacingset#1{\renewcommand{\baselinestretch}%
{#1}\small\normalsize} \spacingset{1}


\if1\blind
{
  \title{Bayesian Nonparametric Adjustment of Confounding}
 \author{\small Chanmin Kim\thanks{
    The authors gratefully acknowledge that this work is supported by the National Research Foundation of Korea
(NRF) grant funded by the Korea government (No. NRF-2020R1F1A1A01048168), US National Institutes of Health (NIHR01ES026217), and US Environmental Protection Agency (EPA 83587201). Its contents are solely the responsibility of the grantee and do not necessarily represent the official views of the supporting agencies. The views expressed in this paper are those of the authors and do not necessarily represent the view or policies of the US Environmental Protection Agency. Further, the US government does not endorse the purchase of any commercial products or services mentioned in the publication. }\hspace{.2cm}\\
   \small Department of Statistics, SungKyunKwan University, Seoul, Korea\\
    \small and \\
    \small Mauricio Tec \\
    \small Department of Statistics and Data Science, The University of Texas, Austin, TX 78712 \\
    \small and \\
    \small Corwin M. Zigler \\
    \small Department of Statistics and Data Science, The University of Texas, Austin, TX 78712 }
    \date{}
  \maketitle
} \fi

\if0\blind
{
  \bigskip
  \bigskip
  \bigskip
  \begin{center}
    {\LARGE\bf Bayesian Nonparametric Adjustment of Confounding}
\end{center}
  \medskip
} \fi

\bigskip
\begin{abstract}
Analysis of observational studies increasingly confronts the challenge of determining which of a possibly high-dimensional set of available covariates are required to satisfy the assumption of ignorable treatment assignment for estimation of causal effects. We propose a Bayesian nonparametric approach that simultaneously 1) prioritizes inclusion of adjustment variables in accordance with existing principles of confounder selection;  2) estimates causal effects in a manner that permits complex relationships among confounders, exposures, and outcomes; and 3) provides causal estimates that account for uncertainty in the nature of confounding.  The proposal relies on specification of multiple Bayesian Additive Regression Trees models, linked together with a common prior distribution that accrues posterior selection probability to covariates on the basis of association with both the exposure and the outcome of interest. A set of extensive simulation studies demonstrates that the proposed method performs well relative to similarly-motivated methodologies in a variety of scenarios. We deploy the method to investigate the causal effect of emissions from coal-fired power plants on ambient air pollution concentrations, where the prospect of confounding due to local and regional meteorological factors introduces uncertainty around the confounding role of a high-dimensional set of measured variables. Ultimately, we show that the proposed method produces more efficient and more consistent results across adjacent years than alternative methods, lending strength to the evidence of the causal relationship between SO2 emissions and ambient particulate pollution.
\end{abstract}

\noindent%
{\it Keywords:}  BART, confounder selection, pollutant emissions, ambient PM2.5
\vfill

\newpage
\spacingset{1.73} 

\section{Introduction}
Advances in statistical methodology have resulted in improved procedures for estimating causal effects from observational studies. Such analyses always rely on the so-called ``no unmeasured confounding'' or conditional ignorability assumption that all confounders are measured and included in the model or adjustment procedure. Ideally, the optimal set of variables would be known based on knowledge of the underlying causal structure of the problem under investigation. However, in many practical settings, the underlying causal structure is not known with certainty, especially when the number of potential confounders is large or the possibility of nonlinearities or interactions in relationships with exposures and outcomes is incompletely understood. As modern analysis settings increasingly leverage large and heterogeneous data sets, researchers are increasingly confronted with the need to choose which of the observed confounders to include in the analysis.

For example, studies of the causal effects of air pollution emissions on ambient air quality routinely entail uncertainty surrounding which of a large set of potential confounders are required for estimating causal effects. We consider the effect of sulfur dioxide (SO$_2$) emissions from coal-fired power plants on the concentration of ambient fine particulate matter (PM$_{2.5}$) during 2013 and 2014 in the United States, representing a key relationship underlying various regulatory strategies. Available data are comprised of emissions from 397 power plants, ambient PM$_{2.5}$ at 12943 ZIP codes across the country, population demographic variables measured at each ZIP code, and seasonal meteorological factors that dictate the formation and transport of PM$_{2.5}$. Of particular note, the regional nature of meteorologic confounding in this context means that not only local conditions but also meteorological variables in surrounding areas should be considered as potential confounding variables, which greatly expands the number of available potential confounders. 


Increasing appreciation that many variable selection methods based on regularized regression (such as the LASSO \citep{tibshirani1996regression} and its variations) are sub optimal for causal effect estimation has motivated a variety of methods for so-called ``confounder selection,'' as a distinct endeavor to the more common setting of variable selection. A common thread of this research is to orient the prioritization of variables to consider both associations with the exposure of interest and the outcome of interest. \cite{wang2012bayesian} proposed a method called Bayesian Adjustment for Confounding (BAC) to conduct variable selection and model averaging on both the exposure and outcome models, linking the two models using unknown nuisance indicators for the inclusion of each potential confounding covariate.  Extensions to BAC have included those to accommodate binary covariates \citep{lefebvre2014extending} and generalized linear models \citep{wang2015accounting}.  BAC can account for uncertainty around confounder selection in the estimation of causal effects; however, it heavily depends on relatively simplistic parametric model assumptions. Wilson and Reich \citep{wilson2014confounder} proposed a similarly-motivated decision-theoretic method that works well for a variety of sample sizes, but poses difficulties in choosing the final solution path for causal estimation. \cite{shortreed2017outcome} proposed the outcome-adaptive LASSO for selecting appropriate covariates for inclusion in propensity score, similarly relying on a parametric linear model and restricted to the case of a binary treatment. \cite{haggstrom2018data} developed a method to learn the causal structure using a probabilistic graphical model and to estimate the causal effect based on the estimated graph. This method is closely linked to existing principles of confounder selection offered in  \cite{vanderweele2011new} and \cite{vanderweele2019principles}, but estimates a single set of covariates and cannot account for uncertainty in confounder selection.

As a related thread of research, decision tree ensembles \citep{freund1999short,breiman2001random} are a powerful tool to predict non-linear dose-response relationships that have received increased attention in the causal inference literature. Among many others, the Bayesian additive regression trees (BART, \cite{chipman2010bart}) method and its modifications have been widely used for obtaining causal estimates. As a key example, the Bayesian causal forest (BCF, \citep{hahn2020bayesian}) model is a variation of the BART that is tailored to estimation of heterogeneous treatment effects. This method is known to outperform the original BART method, especially in situations with small effect sizes, heterogeneous effects, and strong confounding. However, aside from the implicit variable selection in the construction of the component regression trees, the method does not have an explicit feature for confounder selection when there exists a very large set of confounders. 

In this article, we endeavor to unite the objectives of confounder selection procedures such as BAC with the promise of flexible tree-based methods such as BART.  The proposed method orients the flexibility, predictive power, and implicit variable selection involved in BART towards a procedure explicitly defined to select confounders (and other variables) in a manner consistent with principles suggested in, for example, \citep{vanderweele2019principles}, for identifying adjustment variables for estimation of causal effects.  Specifically, the exposure and outcome models are jointly fitted via the BART method with a common prior on a vector of selection probabilities. The joint estimation of the two models generates a selection probability vector that is updated in such a way that more posterior selection probability accrues to variables prioritized in the models for the exposure and/or outcome, based in part on a version of the the sparsity-inducing prior proposed in \cite{linero2018bayesian}. Key benefits of the proposed approach relative to existing methodologies include its avoidance of parametric modeling assumptions, its ability to capture complex relationships and interactions among covariates, exposures, and outcomes, the inherent account of uncertainty in the confounder selection, and the ability to de-emphasize the inclusion of so-called instruments that are associated with the exposure but otherwise unrelated to the outcome of interest.  The usefulness of the method is evidenced - for both binary and continuous exposures - in simulation studies and in an analysis of the causal effects of power plant emissions on ambient air pollution, where the proposed method displays more consistent estimates of causal effects across neighboring years than alternatives.


\section{Motivating Study of the Causal Effects of Power Plant Emissions on Ambient Particulate Pollution}

In the U.S., regulations to limit population exposure to harmful air pollution, mostly from Title IV of the 1990 Amendments to the Clean Air Act, are designed to limit emissions from power plants across the nation. Those emissions lead to poor ambient air quality including an elevated level of fine particulate matter of less than 2.5 $\mu$m (micrometers) in size, also known as PM$_{2.5}$. Thus, there is a growing interest in quantifying the air quality impacts of emissions (or interventions to reduce emissions) from sources such as power plants \citep{zigler2014point,kim2020health}. Coal-fired power plants are of particular interest because large amounts of sulfur dioxide (SO$_2$) are emitted when generating electricity and SO$_2$ is a major precursor to the formation of ambient PM$_{2.5}$. 

\cite{kim2020health} integrate observed data and formal statistical methods with knowledge from the field of atmospheric science to accommodate pollution transport and define the exposure, outcome and confounding variables of interest. Monthly emission of SO$_2$ from electricity generating units (EGUs) are available from the EPA's Air Markets Program Data (AMPD) along with other characteristics of the EGUs including their locations (longitude and latitude). Among a total of 448 coal-fired power plants (1080 EGUs), we use data on the 397 active coal-fired power plants (935 EGUs) during year 2013. For ambient PM$_{2.5}$ outcome data, annual predictions of PM$_{2.5}$ concentrations in grid cells at a resolution of 1 km for year 2013 were obtained from the NASA's socioeconomic data and applications center (https://beta.sedac.ciesin.columbia.edu/data/set/aqdh-pm2-5-concentrations-contiguous-us-1-km-2000-2016) \citep{di2019ensemble}. The gridded data are converted to the zip-code level PM$_{2.5}$ data by averaging over the zip code polygon (see the supplemental material for the R code). The 2010 US Census data (retrieved from the \verb|tidycensus| R package) provide important supporting information on the ZIP codes such as annual measures of total population by age, race, socio-economic status which can be considered as potential confounders. 
Other important potential confounders are weather patterns---which dictate regional differences in the formation and dispersion of ambient air pollution---including temperature, precipitation, cloud cover, planetary boundary layer height, relative humidity, measures of wind speed, wind angles, and the North/South and East/West wind components. To characterize local meteorological conditions in a ZIP code, we assign the value of these variables measured at the centroid of the ZIP code. In addition to local conditions, weather is an inherently regional potential confounding factor that may not be entirely characterized at a given ZIP code by measurements at the centroid \citep{bourikas2013addressing}. To characterize regional meteorological conditions, we also include each meteorological variable (except wind speed and wind angles) measured at 100km displacements in all 8 cardinal directions (N, NE, E, SE, S, SW, W, NW) from the ZIP code centroid.  For all local and displaced weather covariates, we further consider two seasonal values (Summer and Winter) for each weather variable. Note that the account for regional weather introduces 80 additional covariates for each ZIP code, representing each of the 5 meteorological variables measured at the 8 grid cells (one per cardinal direction) located at a 100km radius from that location across the 2 seasons.

To measure each ZIP code's exposure to power plant emissions, we use a recently-developed reduced-complexity atmospheric model, called HYSPLIT Average Dispersion (HyADS), which uses an established Langrangian trajectory model to characterize the impact of power plant emissions \citep{henneman2019characterizing}. The resulting exposure metric represents a weighted sum of SO$_2$ emissions from all coal-fired power plants for the each ZIP code location. See \cite{henneman2019characterizing} for a more detailed explanation and validation of this metric of population exposure to power plant pollution.


\begin{figure}[pt]
\centering
    \includegraphics[width=0.85\textwidth]{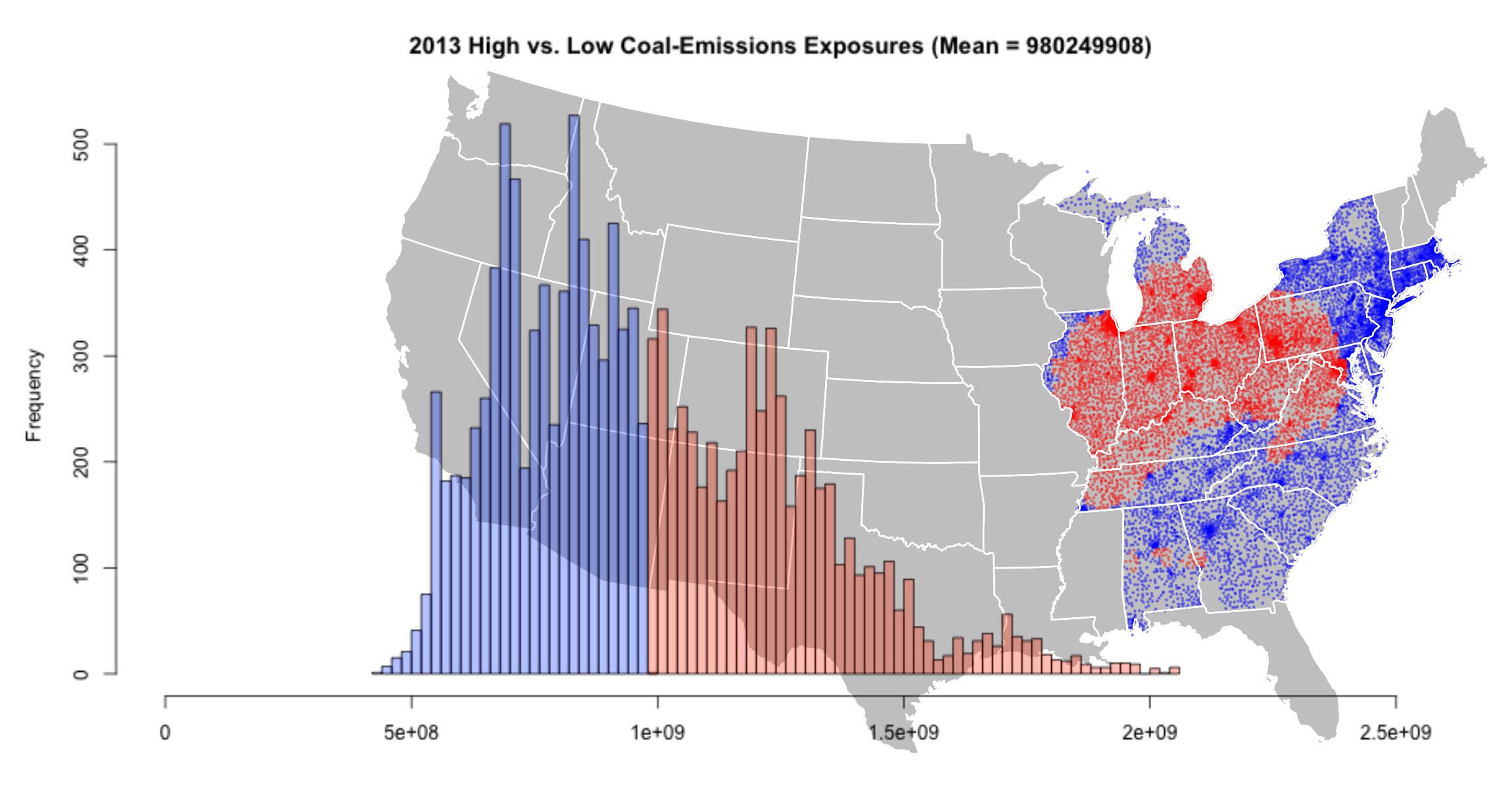}
\caption{\footnotesize The map of the zip code locations: the treated (low HyADS coal-emissions exposure; blue locations) versus the control (high HyADS coal-emissions exposure; red locations) locations. The analysis includes the zip codes located in the Eastern US. The histogram shows frequencies of high or low  HyADS coal-emissions exposures dichotomized at the mean (980249908). }\label{USmap1}
\end{figure}

Owing to the location of most coal-fired power plants in the Midwestern and Eastern US, most of the 23123 zip code locations under consequential coal-emissions exposure are located towards the East. Thus, we restrict the study area of this manuscript to zip-code locations in New York, Pennsylvania, Delaware, New Jersey, Maryland, D.C., Virginia, Massachusetts, Connecticut, Rhode Island, West Virginia, Ohio, Kentucky, Indiana, Illinois, Michigan, Florida, Georgia, South Carolina, North Carolina, Tennessee, and Alabama. We then dichotomize the HyADS exposure ($A$) to define each zip code as either high or low exposed to SO$_2$ emissions based on the mean of the 2013 HyADS coal emissions exposure levels (980,249,908) as in Figure \ref{USmap1}. While we focus discussion on this case of a binary treatment and continuous outcome for illustration, note that extensions other types of exposure and/or outcome data (e.g., a continuous exposure or binary outcome) are available, as will be discussed and illustrated later. Table 1 summarizes potential confounders by the exposure (low HyADS exposure versus high HyADS exposure). While the initial set of potential local and regional confounders included 216 variables, we omit both regional and local variables for surface temperature, convective precipitation, downward shortwave radiation, (north-south) wind component, and (east-west) wind component, and regional variables for wind speed and wind angle to reduce extreme correlation among the set of possible confounders. A total of 104 potential confounders (5 weather variables $\times$ 9 locations $\times$ 2 seasons $+$ wind speed and angle $\times$ 2 seasons $+$ 10 census variables) are available for the analysis, but there is uncertainty around which such variables are required for estimation of causal effects.

\begin{table}[t]
{\renewcommand{\arraystretch}{0.75}
\caption{\label{Data} \footnotesize Summary statistics for selected potential confounders (among 104 potential confounders) and outcome. $\dagger$ Displacements in 8 cardinal directions (N, NE, E, SE, S, SW, W, NW) for a radius of 100km in Summer and Winter are omitted here (i.e., only local weather variables are summarized).}
\centering
\fbox{%
{\scriptsize
\begin{tabular}{lcccccc}
 && \multicolumn{2}{c}{\bf Low HyADS exposure}  & &
 \multicolumn{2}{c}{\bf High HyADS exposure}  \\
 && \multicolumn{2}{c}{A = 1 (n=7216)}  & &
 \multicolumn{2}{c}{A=0 (n=5727)} \\
& & Median & IQR & & Median & IQR\\
\hline
\bf Air Quality and Weather Data$^\dagger$ & & & & & & \\
\hskip1em Ambient PM$_{2.5}$   & & 8.12    &  (7.17, 8.93)    &  &    9.58     & (8.67, 10.16)\\
\hskip1em $\dagger$Temperature at 2m (\texttt{temp}, $^\circ$C, Summer) & & 22.17  &  (20.36, 24.24)     & &   21.64  &    (20.73, 22.89)\\
\hskip1em $\dagger$Temperature at 2m (\texttt{temp}, $^\circ$C, Winter) & & 2.29  &  (-0.41, 6.54)     & &   0.03  &    (-1.13, 2.18)\\
\hskip1em $\dagger$Total Precipitation (\texttt{apcp}, kg/m$^2$, Summer) & & 3.13  &  (2.47, 3.89)     & &   2.34  &    (1.92, 2.81)\\
\hskip1em $\dagger$Total Precipitation (\texttt{apcp}, kg/m$^2$, Winter) & & 2.46  &  (2.14, 3.69)     & &   2.19  &    (1.88, 2.74)\\
\hskip1em $\dagger$Total Cloud Cover (\texttt{tcdc}, \%, Summer) & & 39.40  &  (36.25, 48.98)     & &   38.40  &    (34.92, 41.84)\\
\hskip1em $\dagger$Total Cloud Cover (\texttt{tcdc}, \%, Winter) & & 55.09  &  (51.29, 59.81)     & &   58.74  &    (55.63, 65.52)\\
\hskip1em $\dagger$Planet Boundary Layer Height (\texttt{hpbl}, m, Summer) & & 734.37  &  (658.47, 783.79)     & &   746.96  &    (713.32, 794.77)\\
\hskip1em $\dagger$Planet Boundary Layer Height (\texttt{hpbl}, m, Winter) & & 941.26  &  (825.51, 985.39)     & &   870.94  &    (820.20, 951.08)\\
\hskip1em $\dagger$Relative Humidity (\texttt{rhum}, \%, Summer) & & 79.65  &  (78.46, 80.60)     & &   76.85  &    (74.95, 78.38)\\
\hskip1em $\dagger$Relative Humidity (\texttt{rhum}, \%, Winter) & & 76.09  &  (73.87, 78.28)     & &   76.74  &    (75.44, 79.20)\\
\hskip1em Wind Speed (\texttt{wspd}, m/s, Summer) & & 1.31  &  (1.01, 1.59)     & &   0.72  &    (0.49, 0.93)\\
\hskip1em Wind Speed (\texttt{wspd}, m/s, Winter) & & 2.23  &  (1.33, 2.77)     & &   2.16  &    (1.78, 2.59)\\
\hskip1em Wind Angle (\texttt{phi}, $^\circ$, Summer) & & 241.61  &  (149.83, 254.06)     & &   214.74  &    (183.44, 235.69)\\
\hskip1em Wind Angle (\texttt{phi}, $^\circ$, Winter) & & 278.54  &  (255.65, 293.45)     & &   262.82  &    (255.73, 274.97)\\
\bf Census Data & & & & & & \\
\hskip1em Population (Total) &&  6164 & (2105, 17341)  & &   4118  & (1565, 14293)\\
\hskip1em Population ($<$20 years, \%) &&  25.27 & (22.93, 27.63)  & &   25.95  & (23.76, 28.02)\\
\hskip1em Population (White, \%) &&  89.19 & (70.20, 96.27)  & &   96.28  & (87.62, 97.94)\\
\hskip1em Population (Male, \%) &&  49.08 & (48.08, 50.12)  & &   49.53  & (48.53, 50.57)\\
\hskip1em Population (Female \& Bachelors, \%) &&  7.96 & (5.19, 11.95)  & &   6.74  & (4.51, 9.78)\\
\hskip1em Population (Poverty Status, \%) &&  11.24 & (6.29, 18.21)  & &   10.85  & (6.76, 16.33)\\
\hskip1em Gini Index &&  0.4278 & (0.3940, 0.4674)  & &   0.4120  & (0.3773, 0.4482)\\
\hskip1em Median Income (\$) &&  57991.5 & (44518.3, 79350.8)  & &   56118  & (46604, 69186)\\
\hskip1em Median Age (years) &&  41.3 & (38.2, 44.2)  & &   41.0  & (38.4, 43.5)\\
\hskip1em Housing Units (Urban, \%) &&  47.73 & (0.0, 98.14)  & &   16.72  & (0.0, 87.76)\\
\end{tabular}}
}}\label{tab:data0}
\end{table}

\section{Model}
\subsection{Causal model}
The analysis goal is to quantify the causal effect of an exposure, denoted $A$, on an outcome of interest, denoted $Y$, in a manner that adjusts for observed confounders. Let $\boldsymbol{X} = \{X_1, X_2, \cdots, X_P\}$ be a set of $P$ observed pre-treatment variables that are regarded as potential confounders. In many applications including our motivating power plant investigation, the number of potential confounders is large and we do not have a priori knowledge about which should be adjusted for in the causal estimation, in part due to the possibility of complex inter-relationships among true confounders and their relationships with exposures and outcomes and, in the motivating case, the possibility of confounding due to measurements at other locations.
Recall that for illustration we assume the exposure and outcome variables are binary and continuous, respectively, but generalizations to other types of data are briefly described in Section 3.3.

We formalize the estimation within a potential outcomes framework \citep{rubin1974estimating}. Among a sample of $i=1,2,\ldots,N$ observational units, the \emph{potential outcome} for unit $i$ is defined as $Y_i(a)$ denoting the potential value of the outcome $Y$ that could be observed under the exposure status $A=a$. Then, the target causal estimand $\Delta(a, a^\prime)$ is 
\vspace{-0.5cm}
{\small
\[\Delta(a, a^\prime) = E[Y(a) - Y(a^\prime)],\vspace{-0.5cm}
\]}
which is estimated by the following equation with observed data under the strongly ignorable treatment assignment assumption, that is, $Y_i(1), Y_i(0) \perp A_i | \boldsymbol{X}_i$ \citep{rosenbaum_central_1983}
\vspace{-0.5cm}
{\small
\begin{equation}
\Delta(a, a^\prime; \boldsymbol{x}) = E[Y|A=a, \boldsymbol{X}=\boldsymbol{x} ]- E[Y|A=a^\prime,\boldsymbol{X}=\boldsymbol{x}] \vspace{-0.5cm}
.\label{causal}
\end{equation}
}
This quantity can be marginalized over confounders $\boldsymbol{X}$ to obtain the population effect, $\Delta(a, a^\prime)$. Since it is crucial to meet the strongly ignorable treatment assignment assumption in the estimation, selecting a proper set of confounders to condition upon and finding their inter-relationship in the model $E[Y|A,X]$ are the most important tasks.  We address the second task with an overview of Bayesian Additive Regression Trees (BART) in the next section, returning to the problem of determining the proper set of confounders subsequently.

\subsection{Overview of BART method}
We briefly review the Bayesian additive regression trees method based on \cite{kapelner2013bartmachine}. The general BART model is expressed as follows:
\vspace{-0.5cm}
{\small\[y_i=f(\mathbf{X}_i)+\epsilon_i \approx \sum_{h=1}^H g(\mathbf{X}_i; \mathcal{T}_h, \mathcal{M}_h) + \epsilon_i, \,\,\,\,\,\, \epsilon_i \sim N(0, \sigma^2), \,\, \text{ for } i=1, \cdots, N,\vspace{-0.5cm}
 \]}
where $y_i$ and $\mathbf{X}_i$ are the $i$-th response and vector of predictors, respectively, and $g(\mathbf{X}; \mathcal{T}_h, \mathcal{M}_h)$ is dictated by an estimated tree structure.  Specifically, each of $H$ distinct tree structures is denoted by $\mathcal{T}_h$ ($h=1, \cdots, H$) and the parameters denoting the mean outcome at the terminal nodes of the $h$-th tree are denoted by $\mathcal{M}_h = \{\mu_{h,1}, \cdots, \mu_{h,n_h}\}$ where $n_h$ is the number of terminal nodes of $\mathcal{T}_h$. The function $g(\mathbf{X}; \mathcal{T}_h, \mathcal{M}_h)$ represents $\mu_{h,\eta} \in \mathcal{M}_h$ if $\mathbf{X}$ is associated to the $\eta$-th terminal node in tree $\mathcal{T}_h$.

Each tree structure $\mathcal{T}_h$ contains internal nodes with corresponding ``splitting rule'' (e.g., $X_j > c$) consisting of the ``splitting variable'' $X_j$ and the ``splitting value'' $c$. The tree grows down with these splits until a terminal node is reached where each parameter value $\mu_{t, \eta}$(e.g., leaf value) is assigned. 

The BART method uses ``Bayesian backfitting" \citep{hastie2000bayesian} for a Metropolis-within-Gibbs sampler \citep{geman1993stochastic} where each tree $j$ is fit iteratively through the residual responses (the unexplained responses). In each iteration, a new tree structure is proposed via three distinct tree alterations: (G) GROW, (P) PRUNE, and (C) CHANGE \citep{kapelner2013bartmachine, linero2018bayesian}. \cite{chipman2010bart} originally considered an additional alteration, SWAP, which has been omitted in recent studies.
\begin{itemize}
\item[(G)] Randomly select a terminal node and give birth to two children (i.e., two terminal nodes). This entails random sampling of the predictor $X_j$ and the associated cutpoint $c$ to generate two new children. 
\item[(P)] Randomly select an internal node whose children are both terminal nodes (referred to as a singly internal node) and turn it into a terminal node (i.e., delete two children).
\item[(C)] Randomly select an internal node and change its splitting variable and splitting value based on the prior.
\end{itemize}
In GROW and CHANGE alterations, a new predictor is randomly sampled from a pool of $P$ predictors according to assumed priors. \cite{chipman2010bart} originally proposed a uniform prior, $1/P$, on selection probabilities $s=(s_1, \cdots, s_P)$ of $P$ predictors. Recently, \cite{linero2018bayesian} replaced it with a Dirichlet prior to obtain adaptivity to sparcity, $(s_1, \cdots, s_P) \sim \mathcal{D}(\alpha/P, \cdots, \alpha/P)$.

\subsection{BART confounder selection}
In the spirit of the Bayesian Adjustment for Confounding (BAC) approach, we assume two models for estimating the effect of exposure $A$ on outcome $Y$ along with a pool of high dimensional potential confounders $\mathbf{X}=\{X_1, \cdots, X_P\}$:
\vspace{-0.5cm}
{\small
\[E(A_i|\mathbf{X}_i) = f_a(\mathbf{X}_i), \qquad 
E(Y_i|A_i, \mathbf{X}_i) = f_y(A_i, \mathbf{X}_i)\vspace{-0.5cm}
\]
}
where $f_a(\cdot)$ and $f_y(\cdot)$ are some link functions. In \cite{wang2012bayesian,zigler2014uncertainty,wang2015accounting}, parametric generalized linear models are assumed with vectors of indicators $\mathbf{\alpha}^X$ and $\mathbf{\alpha}^Y$ indicating whether each covariate is selected into the treatment and outcome models, respectively.  High posterior probability of including confounders in the adjustment is accomplished through a joint prior on  $\mathbf{\alpha}^X$ and $\mathbf{\alpha}^Y$ that renders variables selected into the tretment model more likely to be selected into the outcome model (thus prioritizing covariates associated with both $A$ and $Y$). The average causal effect is estimated by Bayesian model averaging (BMA, \cite{raftery1997bayesian}); averaging over all possible outcome models, with posterior weight in the averaging concentrated on the models that include the confounders.

A drawback of the aforementioned methods is they are not able to explore all possible model forms such as nonlinear terms in $f_a(\cdot)$ or/and $f_y(\cdot)$. We propose a new flexible method to account for uncertainty in confounder selection using Bayesian additive regression trees (BART) model that does not depend on parametric model strcutures.  
Using a BART method, we build two models:
\vspace{-0.1cm}
{\small
\begin{eqnarray}
P(A_i=1) &=& \Phi^{-1}\left(\sum_{h=1}^H g_a(\mathbf{X}_i; \mathcal{T}_h, \mathcal{M}_h)\right)\label{model:trt}\\
y_i & = & \sum_{h=1}^{H} g_y(a_i, \mathbf{X}_i; \mathcal{T}_{h}^\prime, \mathcal{M}_{h}^\prime) + \epsilon_i \,\,\,\,\,\, \text{ where } \,\, \epsilon_i \sim N(0, \sigma^2)
\vspace{-0.1cm}
\label{model:out}
\end{eqnarray}
}
for subjects $i=1,\cdots, N$ where $\Phi^{-1}$ is the inverse standard normal cdf. Note that it is straightforward to accommodate a continuous exposure by replacing Eq (\ref{model:trt}) with
\vspace{-0.2cm}
{\small
\[A_i  =  \sum_{h=1}^{H} g_a(\mathbf{X}_i; \mathcal{T}_{h}, \mathcal{M}_{h}) + \omega_i \,\,\,\,\,\, \text{ where } \,\, \omega_i \sim N(0, \tau^2).\vspace{-0.3cm}
\] 
}
In a manner analogous to BAC, the exposure and outcome models are fitted jointly to assign posterior weight to models that emphasize the ``right'' confounders.  Instead of introducing explicit covariate inclusions parameters (e.g., $\alpha^X, \alpha^Y$ in BAC), the exposure and outcome models are linked via common prior on selection probabilities $\boldsymbol{s}=(s_1 \cdots, s_P)$ of $P$ potential confounders in GROW and CHANGE alterations. Specifically, we use a common sparsity-inducing Dirichlet prior $(s_1, \cdots, s_P) \sim \mathcal{D}(\alpha/P, \cdots, \alpha/P)$ which gives a conjugate Gibbs-sampling update,
{\small
$(s_1, \cdots, s_P) \sim \mathcal{D}(\alpha/P+m_1+n_1, 2\alpha/P+m_2+n_2, \cdots, 2\alpha/P+m_P+n_P)$,
}
where $m_j$ and $n_j$ denote the number of splits on confounder $X_j$ in Model (\ref{model:trt}) and Model (\ref{model:out}), respectively. This conjugate update is based on the following likelihood of $\boldsymbol{s}$ 
\vspace{-0.3cm}
{\small
\begin{equation}
\prod_{h=1}^H \left\{\prod_{b\in \mathcal{T}_h} s_{j_b} \prod_{c\in \mathcal{T}^\prime_h} s_{j_c}\right\}, \vspace{-0.3cm}
\label{likelihood}
\end{equation}
}
where $b$ and $c$ denote nodes of trees $\mathcal{T}_h$ and $\mathcal{T}^\prime_h$, respectively, and $j_b$ and $j_c$ denote the predictors used to split nodes $b$ and $c$, respectively.
This likelihood is correct under a certain assumption (Assumption 2.2 in \cite{linero2018bayesian}) which is acceptable whenever the number of unique values of each confounder is sufficiently large or the trees are typically very shallow, which is often reasonable. As long as continuous covariates (or at least discrete variables with many categories) are used with a large $H$ (i.e., the number of trees), this assumption generally holds.


If a given covariate, $X_j$, is used often as a splitting variable in either the model for $A$ or the model for $Y$, the model will accumulate posterior mass on the selection probability $s_j$ through larger values of $m_j$ or $n_j$, with the most selection probability accumulating for variables that are important in the prediction of {\it both} $A$ and $Y$ (which will accumulate larger values of $m_j$ and $n_j$). Thus, selection probabilities will tend to favor the $X_j$ that are associated with $A$, associated with $Y$, or associated with both $A$ and $Y$.  The variables ultimately used for effect estimation in the model for $Y$ will be those that are proposed for splitting via this prior and accepted in the updating step of the model for $Y$, which will further prioritize variables associated with $Y$.  We discuss this confounder prioritization in detail in Section 5.

\subsubsection{Separate models for each exposure arm}
However, further caution is warranted in fitting Model (3). Since it includes $A$, which is not a potential confounder but plays a crucial role in estimating the causal effect using the fitted observed data model, the model should be sure to avoid a low selection probability for $A$. The inclusion of $A$ in model (3) also makes the dimension of $\boldsymbol{s}$ in each model different, which complicates the conjugate updating of $\boldsymbol{s}$.

Thus, a modified version of Model (3) is proposed for each $A_i=0, 1$,
\vspace{-0.3cm}
{\small
\begin{eqnarray}
y_i & = & \sum_{h=1}^{H} g_y^0(\mathbf{X}_i; \mathcal{T}_{h}^0, \mathcal{M}_{h}^0) + \epsilon_i^0, \,\,\,\,\,\, \epsilon_i^0 \sim N(0, \sigma^2_0)\,\,\,\, \text{ for } i \in I_0\\
y_i & = & \sum_{h=1}^{H} g_y^1(\mathbf{X}_i; \mathcal{T}_{h}^1, \mathcal{M}_{h}^1) + \epsilon_i^1, \,\,\,\,\,\,\epsilon_i^1 \sim N(0, \sigma^2_1)\,\,\,\, \text{ for } i \in I_1 \vspace{-0.5cm}
\end{eqnarray}
}
where $I_0$ and $I_1$ denote two sets of observations under $A=0$ and $A=1$, respectively. Practically, three models, Models (2), (5) and (6), are fitted with a common prior on the selection probabilities $\boldsymbol{s} = (s_1, \cdots, s_P)$ with a conjugate sampling update
\vspace{-0.5cm}
{\small
\begin{equation}
(s_1, \cdots, s_P) \sim \mathcal{D}(\alpha/P+m_1+n_1^0+n_1^1, \cdots, \alpha/P+m_P+n_P^0+n_P^1 ),\label{selection.prob}\vspace{-0.5cm}
\end{equation}
}
where $n_j^0$ and $n_j^1$ are the number of splits on confounder $X_j$ in two outcome models (5, 6).

\subsubsection{Single marginal model}
Early work in \cite{hill2011bayesian} pointed out a key caution with using separate outcome models for each exposure level. If there exist parts of the covariate space with little or no observed representation from both exposure groups (which is referred to as lack of common support), estimating two separate outcome models for exposures $A=0$ and $A=1$ may  produce highly biased estimates. \cite{hahn2020bayesian} also advocated using a single outcome model and proposed the BCF which will be discussed in detail later.

Using a single outcome model for the confounder selection case introduces a key challenge for the confounder selection case.  Note that the outcome model has an additional selection probability parameter $s_0$ that is tied to the exposure variable in the vector of selection probabilities $\boldsymbol{s} = (s_0, s_1, \cdots, s_P)$ while the exposure model only uses the subset $\boldsymbol{s}^\prime = (s_1^\prime, s_2^\prime, \cdots, s_P^\prime)$ in the selection of $P$ predictors. Thus, it is not straightforward to specify a common prior on the vectors of selection probabilities $\boldsymbol{s}$ and $\boldsymbol{s}^\prime$.

We link two vectors using a common prior by reparameterizing $\boldsymbol{s}^\prime = (s_1^\prime, s_2^\prime, \cdots, s_P^\prime)=(s_1/(1-s_0), s_2/(1-s_0), \cdots, s_P/(1-s_0))$ based upon neutrality of the Dirichlet distribution. With the Dirichlet distribution $\mathcal{D}(\alpha/P, \alpha/P, \cdots, \alpha/P)$ for the common prior for $\boldsymbol{s} = (s_0, s_1, \cdots, s_P)$, the update of $\boldsymbol{s}$ is based on the following likelihood$\times$prior ($\mathcal{Q}$)
\vspace{-0.3cm}
{\small
\[\mathcal{Q}=\prod_{h=1}^H \left\{\prod_{b\in \mathcal{T}_h} \frac{s_{j_b}}{1-s_0} \prod_{c\in \mathcal{T}^\prime_h} s_{j_c}\right\} \prod_{j=0}^P s_j^{\alpha/P-1}, 
\vspace{-0.3cm}\]
}
which is valid under the same assumption used in Eq. (\ref{likelihood}). This is equivalent to 
\vspace{-0.3cm}
{\small
\[\mathcal{Q} = \left(\frac{1}{1-s_0 }\right)^{\sum_{j=1}^P m_j} s_0^{n_0+\alpha/P-1}s_1^{m_1+n_1+\alpha/P-1}s_2^{m_2+n_2+\alpha/P-1} \cdots s_P^{m_P+n_P+\alpha/P-1},\vspace{-0.3cm}\]
}
where $m_j$ and $n_j$ denote the numbers of splits on confounder $X_j$ in the exposure and outcome models, respectively. Since we no longer entertain an efficient conjugate update, we use the Metropolis-Hastings algorithm to update $\boldsymbol{s}$ with a Dirichlet proposal distribution based on $\mathcal{Q}$, that is $\mathcal{D}(n_0+\alpha/P, m_1+n_1+\alpha/P,m_2+n_2+\alpha/P, \cdots, m_J+n_J+\alpha/P)$, through the following acceptance ratio
\vspace{-0.4cm}
{\small
\[P_{\text{AR}}(\boldsymbol{s} \rightarrow \boldsymbol{s}^{\text{new}}) = \text{min} \left\{1, \left(\frac{\sum_{j=1}^P s_j}{\sum_{j=1}^Ps_j^{\text{new}}}\right)^{\sum_{j=1}^P m_j}  \right\},\vspace{-0.4cm}\]
}
where $\boldsymbol{s}^\text{new}$ is a proposed vector.

One potential drawback of this approach is that the exposure variable is now a part of the variables to be selected based on $\boldsymbol{s}$. To avoid a situation that the exposure variable is rarely used in the tree structure, we can consider a modified Dirichlet proposal distribution $\mathcal{D}(n_0+c+\alpha/P, m_1+n_1+\alpha/P,m_2+n_2+\alpha/P, \cdots, m_J+n_J+\alpha/P)$ where $c$ is some positive number. Since other probabilities $(s_1, \cdots, s_P)$ are subject to additional powers based on the numbers of splits (i.e., $m_j$'s) used in the exposure models, the selection probability $s_0$ may be underweighted in the original $\mathcal{Q}$. In our simulation and application studies, we use $c = n_0$, which results in $s_0^{2n_0 + \alpha/P-1}$ component in the Dirichlet proposal distribution.

\section{Estimation}
\subsection{Posterior computation for the BART models}\label{sec:posterior}
To draw posterior samples from $P(\mathcal{T}_1^\prime, \cdots, \mathcal{T}_H^\prime, \mathcal{M}_1^\prime, \cdots, \mathcal{M}_H^\prime, \sigma^2 | \boldsymbol{D})$ for the outcome model (3) (or (5) and (6)), we use ``Bayesian backfitting" \citep{hastie2000bayesian} for a Metropolis-within-Gibbs sampler where each tree $\mathcal{T}_h^\prime$ is fit iteratively via the residual responses:
\vspace{-0.2cm}
{\small
\[R_{i,-j} = y_i - \sum_{h\neq j} g_y(\mathbf{X}_i; \mathcal{T}_h^\prime, \mathcal{M}_h^\prime), \,\, \text{ for } i=1, \cdots, N.\vspace{-0.2cm}\]
}
For each tree $j$, a new tree structure $\mathcal{T}_j^\prime$ is proposed from the full conditional $[\mathcal{T}_j^\prime | R_{1, -j}, \cdots, R_{n, -j}, \sigma_2]$ (i.e., grow, prune or change alterations) and the parameter within the tree updated through the full conditional $[\mathcal{M}_j^\prime | \mathcal{T}_j^\prime, R_{1, -j}, \cdots, R_{n, -j}, \sigma_2]$ with acceptance ratios for three alteration steps. \cite{kapelner2013bartmachine} derive the exact forms of the acceptance ratios for three alteration steps which are provided in the appendix. Given tree $j$, we draw samples from $P(\mathcal{M}_j^\prime | \mathcal{T}_j^\prime)$ based on the prior $\mu \sim N(\mu_\mu/H, \sigma^2_\mu)$ on each of the leaf parameters $\mathcal{M}_j^\prime:=\{\mu_1, \mu_2, \cdots, \mu_b\}$ where $b$ is the number of terminal nodes in tree $\mathcal{T}_j^\prime$. Here, $\mu_\mu$ is set to the range center of the outcome and $\sigma^2_\mu$ is empirically set to satisfy $H \mu_\mu - 2 \sqrt{H} \sigma_\mu = y_{\text{min}}$ and $H \mu_\mu + 2 \sqrt{H} \sigma_\mu = y_{\text{max}}$ \citep{kapelner2013bartmachine}. For the $\eta$-th terminal node in tree $\mathcal{T}_j^\prime$, we draw sample $\mu_\eta$ based on  
\vspace{-0.2cm}
{\small
\[\mu_\eta \sim N\left(\frac{1}{1/\sigma_\mu^2+n_\eta/\sigma^2} \left(\frac{\mu_\mu/H}{\sigma^2_\mu}+\frac{\sum_{i\in I_\eta} R_{i, -j}}{\sigma^2}\right), \left(\frac{1}{\sigma_\mu^2}+\frac{n_\eta}{\sigma^2}\right)^{-1}\right),\vspace{-0.2cm}\]
}
where $I_\eta$ and $n_\eta$ denotes observation indices corresponding to the $\eta$th terminal node and the number of observations in that node, respectively. 
If the separate model scheme is used, we draw samples from $P(\mathcal{T}_1^1, \cdots, \mathcal{T}_H^1, \mathcal{M}_1^1, \cdots, \mathcal{M}_H^1, \sigma^2_1 | \boldsymbol{D})$ and $P(\mathcal{T}_1^0, \cdots, \mathcal{T}_H^0, \mathcal{M}_1^0, \cdots, \mathcal{M}_H^0, \sigma^2_0 | \boldsymbol{D})$ using the following backfitting steps:
\vspace{-0.2cm}
{\small
\[
R_{i,-j} = y_i - \sum_{h\neq j} g_y^0(\mathbf{X}_i; \mathcal{T}_h^0, \mathcal{M}_h^0) \,\, \text{ for } i \in I_0, \quad\quad
R_{i,-j} = y_i - \sum_{h\neq j} g_y^1(\mathbf{X}_i; \mathcal{T}_h^1, \mathcal{M}_h^1) \,\, \text{ for } i \in I_1,
\vspace{-0.2cm}
\]
}
where $I_0$ and $I_1$ denote two sets of observations under $A=0$ and $A=1$, respectively. 

To draw posterior samples from $P(\mathcal{T}_1, \cdots, \mathcal{T}_H, \mathcal{M}_1, \cdots, \mathcal{M}_H | \boldsymbol{D})$ for the binary exposure model (2), we introduce latent variable $Z$ as follows
{\small
\[ Z_i \sim \left\{ \begin{array}{ll}
         N\left(\sum_{h=1}^H g_a(\boldsymbol{X}_i; \mathcal{T}_h, \mathcal{M}_h), 1\right)I_{(Z_i  > 0)} & \mbox{for $A_i =1 $};\\
        N\left(\sum_{h=1}^H g_a(\boldsymbol{X}_i; \mathcal{T}_h, \mathcal{M}_h), 1\right)I_{(Z_i  \leq 0)} & \mbox{for $A_i = 0 $}\end{array} \right. \vspace{-0.2cm} \] 
        }
for $i=1, \cdots, n$ and apply the general BART model for continuous data to the latent variable $Z$. If $A$ is continuous, the updating is analogous to Bayesian backfitting described for model (3). After updating all tree structures and the corresponding parameters, we update the variance parameters ($\sigma^2$ in the outcome model (3)) based on the Gibbs sampler and the final residuals:
\vspace{-0.3cm}
{\small
\[\sigma^2 \sim \text{Inv.Gamma}\left(a_\sigma+\frac{N}{2}, b_\sigma+\frac{1}{2}\left\{\sum_{i=1}^{N} \left(y_i - \sum_{h=1}^H g_y(\boldsymbol{X}_i; \mathcal{T}_h^\prime, \mathcal{M}_h^\prime)\right) \right\}\right),\vspace{-0.3cm}\]
}
where we set $a_\sigma=b_\sigma=3$ throughout the manuscript as suggested in \cite{chipman2010bart}. The separate model needs to update two variance parameters for the two outcome models
\vspace{-0.2cm}
{\small
\begin{eqnarray*}
\sigma^2_0 &\sim& \text{Inv.Gamma}\left(a_\sigma+\frac{N}{2}, b_\sigma+\frac{1}{2}\left\{\sum_{i \in I_0} \left(y_i - \sum_{h=1}^H g_y^0(\boldsymbol{X}_i; \mathcal{T}_h^0, \mathcal{M}_h^0)\right) \right\}\right),\\
\sigma^2_1 &\sim& \text{Inv.Gamma}\left(a_\sigma+\frac{N}{2}, b_\sigma+\frac{1}{2}\left\{\sum_{i \in I_1} \left(y_i - \sum_{h=1}^H g_y^1(\boldsymbol{X}_i; \mathcal{T}_h^1, \mathcal{M}_h^1)\right) \right\}\right).\vspace{-0.5cm}
\end{eqnarray*}
}
If $A$ is continuous, the variance parameter for the exposure model is updated analogously. Then, we update parameter $\alpha$ in the prior distribution of selection probabilities $\boldsymbol{s} \sim \mathcal{D}\left(\frac{\alpha}{P}, \cdots, \frac{\alpha}{P}\right)$. As suggested in \cite{linero2018bayesian}, we select a prior with the form $\frac{\alpha}{\alpha+P} \sim \text{Beta}(a_0, b_0)$ where $a_0 = 0.5$, $b_0=1$ and update the parameter using the Metropolis-Hastings algorithm. Finally, for the marginal model scheme, we use the Metropolis-Hastings algorithm to update the vector of selection probabilities $\boldsymbol{s}$ through the acceptance ratio
\vspace{-0.3cm}
{\small
\[P_{\text{AR}}(\boldsymbol{s} \rightarrow \boldsymbol{s}^{\text{new}}) = \text{min} \left\{1, \left(\frac{1-\sum_{j=1}^P s_j}{1-\sum_{j=1}^Ps_j^{\text{new}}}\right)^{\sum_{j=1}^J m_j}  \right\}.\vspace{-0.3cm}\]

}
For the separate model approach, $\boldsymbol{s}$ is updated with a conjugate sampling update as in Eq.(7). The code implemented to run the MCMC algorithm is available at http://github.com/lit777/bart-cs.

\subsection{Estimation of the causal effect}\label{sec:estimation}
To estimate the target causal estimand $\Delta(a,a^\prime)$ in Eq. (\ref{causal}) for binary exposure (i.e., $a=1,a^
\prime=0$), we use the posterior samples drawn from Section \ref{sec:posterior}. For the separate model, we estimate $\Delta(1,0)$ through the following equation
\vspace{-0.3cm}
{\small
\begin{equation} \hat{\Delta}(1,0)=\frac{1}{N}\sum_{i=1}^N\left[\frac{1}{R} \sum_{r=1}^R \left\{\sum_{h=1}^{H} g_y^{1,(r)}(\mathbf{X}_i; \mathcal{T}_{h}^1, \mathcal{M}_{h}^1) - \sum_{h=1}^{H} g_y^{0,(r)}(\mathbf{X}_i; \mathcal{T}_{h}^0, \mathcal{M}_{h}^0)\right\}\right], \label{eq:causal_estimate}
\vspace{-0.3cm}
\end{equation}
}
where $g_y^{1,(r)}$ and $g_y^{0,(r)}$ are the $r$-th posterior samples for the models (5) and (6). For the marginal model, we evaluate the following equation instead
\vspace{-0.3cm}
{\small
\begin{equation} \hat{\Delta}(1,0)=\frac{1}{N}\sum_{i=1}^N\left[\frac{1}{R} \sum_{r=1}^R \left\{\sum_{h=1}^{H} g_y^{(r)}(1,\mathbf{X}_i; \mathcal{T}_{h}^\prime, \mathcal{M}_{h}^\prime) - \sum_{h=1}^{H} g_y^{(r)}(0,\mathbf{X}_i; \mathcal{T}_{h}^\prime, \mathcal{M}_{h}^\prime)\right\}\right],\label{eq:causal_estimate_marginal}\vspace{-0.3cm}
\end{equation}
}
where $g_y^{(r)}$ is the $r$-th posterior samples for the model (\ref{model:out}). Like BAC, our approach does not require a separate Bayesian model averaging (BMA) step to estimate the causal effects averaged across all possible outcome models. Since each posterior samples of $g_y^{a,(r)}$ or $g_y^{(r)}$ only include selected confounders in their nodes, averaging over posterior samples of $g_y^{a,(r)}$ and $g_y^{(r)}$ suffices for estimation of causal effects averaged over the different model specifications, provided that the outcome model includes at least the minimal set of confounders required to satisfy the assumption of ignorability.  We elaborate on the methods ability to ensure this in the following section.

\section{Confounder Prioritization with BART Confounder Selection}



The proposed method entails two features for prioritizing adjustment variables for estimation of causal effects: the splitting prior and the posterior updating of the outcome model in (\ref{model:out}) (or models in (5) and (6)).  To illustrate, we focus on the the separate model in Section 3.3.1 involving models (5) and (6). Based on the updated selection probabilities in (\ref{selection.prob}), we can show that
\vspace{-0.3cm}
\begin{equation}
\scriptsize
\text{E}(s_j) = \frac{\alpha/P + m_j+n_j^0 + n_j^1}{\alpha + M + N^0 + N^1},\quad \text{Var}(s_j) = \frac{(\alpha/P + m_j+n_j^0 + n_j^1)(\alpha + M + N^0 + N^1-(\alpha/P + m_j+n_j^0 + n_j^1))}{(\alpha + M + N^0 + N^1)^2(\alpha + M + N^0 + N^1+1)},\vspace{-0.3cm}\label{ex.var}
\end{equation}
where $M=\sum_{i=1}^P m_i, N^0 = \sum_{i=1}^P n_i^0$ and $N^1 = \sum_{i=1}^P n_i^1$. 
Denote by $\mathcal{X}_A$ and $\mathcal{X}_Y$ the sets of covariates in the true exposure model and the true outcome model, respectively, which would contain as a subset the minimal set of confounders required to satisfy ignorability, as well as some additional variables. With the sparsity inducing Dirichlet prior, $X_j$ is rarely used as a splitting variable if the $j$th predictor does not improve the fit of either the exposure model or the outcome model \citep{linero2018bayesian}. Given fixed $M-m_j, N^0 - n_j^0, N^1-n_j^1$, if $X_j \notin \mathcal{X}_A \cup \mathcal{X}_Y$, then the total number of splits using $X_j$ ($m_j + n_j^0 + n_j^1$) is sufficiently smaller than $(M+N^1+N^0)$ and we can assume $\frac{m_j + n_j^0 + n_j^1}{M+N^1+N^0} \approx 0$. Thus, from Eq. (\ref{ex.var}) it is easy to see that the posterior distribution of $s_j$ converges to a degenerate measure at 0 as $\alpha/P \downarrow 0$ (i.e., the number of predictors is huge relatively to $\alpha$ value). That is, among the selection probabilities $\boldsymbol{s} = (s_1, \cdots, s_P)$, $s_j$'s will have values approaching 0 if the corresponding predictors $X_j$'s are neither in $\mathcal{X}_A$ nor in $\mathcal{X}_Y$ when the number of predictors is large relative to $\alpha$. Those covariates with 0 selection probabilities are not available as splitting variables in tree alteration steps and will not enter into the models (2), (5) and (6) with probability 1. If $X_j$ is a predictor in either $\mathcal{X}_A$ or $\mathcal{X}_Y$ or both, then $s_j$ will have non-zero posterior probability, as at least one of $m_j, n^0_j, n^1_j$ will grow to produce $\frac{m_j + n_j^0 + n_j^1}{M+N^1+N^0} > 0$.  Thus, the proposed prior ensures that the covariates available for tree construction meet the ``disjunctive cause criterion" \citep{vanderweele2011new}, where the set of predictors controlled for is a cause of the exposure, or the outcome, or both. In general, this criterion is a good alternative to two other approaches \citep{vanderweele2019principles}: (a) the pre-exposure criterion to control for any predictor that is prior to the exposure (which would pertain to all pre-exposure predictors including every $X_j \in \mathcal{X}_A \cup \mathcal{X}_Y$); and (b) the common cause criterion to adjust for all pre-exposure covariates that are common causes of exposure and outcome (which would pertain to $X_j \in (\mathcal{X}_A \cap \mathcal{X}_Y$) only). 

However, it is also known that, inclusion of an ``instrumental variable'' which affects the outcome only through its association with the exposure variable (i.e., $X_j \in \mathcal{X}_A$ but $X_j \notin \mathcal{X}_Y$), can increase the variance of treatment effect estimates, even if it does not introduce bias in estimation of the causal effect.  What's more, it is known that an unmeasured confounder can amplify bias if adjustment is made for the instrumental variable (see Figure S1 in the supplementary material), which is called ``Z-bias'' \citep{ding2017instrumental}. Thus, a better alternative to all of the aforementioned criteria is to exclude from the ``disjunctive cause criterion'' subset any variables exhibiting association with $A$ without exhibiting an association with $Y$, known as the ``disjunctive cause criterion without instruments'' \citep{vanderweele2019principles}.  As shown previously, the proposed joint splitting prior prioritizes the ``disjunctive cause criterion,'' where $s_j$ can accumulate nonzero mass for instrumental variables through large values of $m_j$ owing to many accepted splits in the model for $A$. However, in the tree alteration steps of the outcome models, even if an instrument $X_j$ with $s_j > 0$ is randomly sampled from the pool of covariates and suggested for a next splitting variable, its acceptance with the Metropolis-Hastings step for the next tree structure will relate to whether the proposed split results in improved prediction of $Y$; an instrumental variable not directly related to the outcome will rarely be accepted for a tree split in the model used for estimating causal effects. Thus, the combination of the joint splitting prior and the posterior updating steps for the outcome model renders our method closely linked to the ``disjunctive cause criterion without instruments" criterion. 

The method's adherence to the above confounder selection principles implies that posterior estimates of the causal effects with the expressions in (\ref{eq:causal_estimate}) or (\ref{eq:causal_estimate_marginal}) will be based on the relevant variables for unbiased effect estimation. To illustrate, let $\mathcal{X}_\cap=(\mathcal{X}_A \cap \mathcal{X}_Y)$ be the set of covariates that satisfy the common cause criterion, and $\mathcal{X}_{\star}=(\mathcal{X}_A \cap \mathcal{X}_Y) \cup \mathcal{X}_Y$ be the set of covariates that satisfy the disjunctive cause criterion without instruments.  The set of variables in $\mathcal{X}_{\cap}$ represents the minimal set of confounders required to satisfy the ignorability assumption, and will suffice to control for confounding for the effect of the exposure on the outcome in the absence of unmeasured confounders. While $\mathcal{X}_{\cap}$ represents a minimal set, the set $\mathcal{X}_\star$ is preferred. Denote $\mathcal{I}$, $\mathcal{I}_\cap$ and $\mathcal{I_\star}$ the set of all possible covariate configurations, a subset of covariate configurations that include at least all the covariates in $\mathcal{X}_\cap$ (and possibly others) as a splitting variable at least once, and a subset of covariate configurations that use all the covariates in $\mathcal{X}_{\star}$ as a splitting variable at least once, respectively. Note that $\mathcal{I}_\star \subset \mathcal{I}_\cap \subset \mathcal{I}$. Denote $R$, $R_\cap$ and $R_\star$ the corresponding posterior samples from models in $\mathcal{I}$, $\mathcal{I}_\cap$ and $\mathcal{I_\star}$, respectively. Then, the posterior mean estimate of the target causal estimand in (\ref{eq:causal_estimate}) can be decomposed into two parts as follows
\vspace{-0.3cm}
{\footnotesize
\begin{eqnarray}
\hat{\Delta}(1,0) & = & \frac{1}{|R_{\cap}|} \sum_{r\in R_{\cap}} \left[\frac{1}{N}\sum_{i=1}^N\left\{\sum_{h=1}^{H} g_y^{1,(r)}(\mathbf{X}_i; \mathcal{T}_{h}^1, \mathcal{M}_{h}^1) - \sum_{h=1}^{H} g_y^{0,(r)}(\mathbf{X}_i; \mathcal{T}_{h}^0, \mathcal{M}_{h}^0)\right\}\right] \nonumber\\
&  & + \,\,\frac{1}{|R|-|R_{\cap}|} \sum_{r\in R \setminus R_{\cap}} \left[\frac{1}{N}\sum_{i=1}^N\left\{\sum_{h=1}^{H} g_y^{1,(r)}(\mathbf{X}_i; \mathcal{T}_{h}^1, \mathcal{M}_{h}^1) - \sum_{h=1}^{H} g_y^{0,(r)}(\mathbf{X}_i; \mathcal{T}_{h}^0, \mathcal{M}_{h}^0)\right\}\right], \label{decomp}\vspace{-0.3cm}
\end{eqnarray}
}
where the first term is the sum over models that include the set of confounders in $\mathcal{X}_{\cap}$ and the second term is the remaining models that are visited in the posterior distribution but never split on at least one common cause. In the absence of unmeasured confounding, the first term represents all possible models that produce unbiased estimates of causal effects, as all these models include the necessary confounders.  The second term includes estimates that would be biased with respect to the causal effect, as they omit at least one confounding variable. Note that we can further decompose the first term in Eq.(\ref{decomp}) and rewrite the posterior mean estimate of the causal effect as follows
\vspace{-0.3cm}
{\footnotesize
\begin{eqnarray}
\hat{\Delta}(1,0) & = & \frac{1}{|R_{\star}|} \sum_{r\in R_{\star}} \left[\frac{1}{N}\sum_{i=1}^N\left\{\sum_{h=1}^{H} g_y^{1,(r)}(\mathbf{X}_i; \mathcal{T}_{h}^1, \mathcal{M}_{h}^1) - \sum_{h=1}^{H} g_y^{0,(r)}(\mathbf{X}_i; \mathcal{T}_{h}^0, \mathcal{M}_{h}^0)\right\}\right] \nonumber\\
&  & + \,\,\frac{1}{|R_\cap|-|R_{\star}|} \sum_{r\in R_{\cap} \setminus R_{\star}} \left[\frac{1}{N}\sum_{i=1}^N\left\{\sum_{h=1}^{H} g_y^{1,(r)}(\mathbf{X}_i; \mathcal{T}_{h}^1, \mathcal{M}_{h}^1) - \sum_{h=1}^{H} g_y^{0,(r)}(\mathbf{X}_i; \mathcal{T}_{h}^0, \mathcal{M}_{h}^0)\right\}\right]\nonumber\\
&  & + \,\,\frac{1}{|R|-|R_{\cap}|} \sum_{r\in R \setminus R_{\cap}} \left[\frac{1}{N}\sum_{i=1}^N\left\{\sum_{h=1}^{H} g_y^{1,(r)}(\mathbf{X}_i; \mathcal{T}_{h}^1, \mathcal{M}_{h}^1) - \sum_{h=1}^{H} g_y^{0,(r)}(\mathbf{X}_i; \mathcal{T}_{h}^0, \mathcal{M}_{h}^0)\right\}\right],\vspace{-0.3cm}\label{decomp1}
\end{eqnarray}
}
where the first two terms of summations represent models that will provide unbiased estimation of the causal effect; the first is the sum over models that include the set of confounders in $\mathcal{X}_\star$, the second term is the sum over models that include  the confounders in $\mathcal{X}_{\cap}$. The third term represents models that omit at least one confounder, and will not provide unbiased estimation of the causal effect.

The BART confounder selection procedure proposed above is designed precisely to allocate posterior support to models in the first two terms of expression (\ref{decomp1}) (i.e., large $|R_{\cap}|$ and $|R_{\star}|$ relative to $|R|$), particularly the first.  The joint splitting prior will ensure that splits are proposed in accordance with models in the first two terms, the posterior updates of the model in (\ref{model:out}) (or (5) and (6)) will be accepted to focus posterior weight on the first term.  We illustrate this with simulated data in Section \ref{sec:sim}.


\section{Simulation}\label{sec:sim}

We test our model performance based on a toy example with $N=300$ ($N=500$ for the last scenario) observations. In six different scenarios, 100 potential confounders ($X_1-X_{100}$) are independently generated from $N(0,1)$ where only 5 of them ($X_1-X_5$) are true confounders:
\begin{itemize}
\item Scenario 1: $A$ model contains only true confounders ($X_1 - X_5$); $Y$ model contains true confounders ($X_1 - X_5$) and predictors ($X_6-X_7$).
\item Scenario 2: $A$ model contains only true confounders ($X_1 - X_5$); $Y$ model contains true confounders ($X_1 - X_5$).
\item Scenario 3: $A$ model contains true confounders ($X_1 - X_5$) and `instrumental variables' ($X_6 - X_7$); $Y$ model contains true confounders ($X_1 - X_5$).
\item Scenario 4: $A$ model contains only true confounders ($X_1 - X_5$); $Y$ model contains true confounders ($X_1 - X_5$) and predictors ($X_6-X_{17}$).
\item Scenario 5: $A$ model contains only true confounders ($X_1 - X_5$); $Y$ model contains true confounders ($X_1 - X_5$) and predictors ($X_6-X_{17}$). The outcome and exposure models are the same in Scenario 4, but the sample size is $N=500$.
\item Scenario 6: the same settings in Scenario 4 except the coefficients of the true confounders in the $Y$ model are relatively smaller.
\end{itemize}
See the supplementary materials for the detailed specification.
Table S1 in the supplementary materials summarizes the scenarios in terms of covariates used in $Y$ (outcome) and $A$ (exposure) models. In each scenario, the first two variables are included in the models through non-linear functions $h_1(x) = (-1)^{I(x <0)}$ and $h_2(x) = (-1)^{I(x \geq 0)}$ and the remaining covariates are also added either with an absolute function or with an interaction term. We generate $m=200$ replicates under each scenario. For our proposed approach, we consider both methods discussed in Section 3, the separate and marginal models. The MCMC chain runs for 25,000 iterations and the first half is discarded as burn-in. To reduce autocorrelation among samples, the thinning interval is set to 10.

\subsection{Alternative methods for comparison}
To compare our model performance to others, we consider five similarly-motivated comparison methods: (a) Bayesian causal forest (BCF) from \cite{hahn2020bayesian}; (b) Bayesian adjustment of confounding (BAC) from \cite{wang2015accounting}; (c) inverse probability weighting with generalized boosted regression for propensity scores (twang from \cite{ridgeway2008twang}); (d) Markov/Bayesian network confounder selection (CovSelHigh) from \cite{haggstrom2018data}; and (e) Bayesian penalized credible region (BayesPen) from \cite{wilson2014confounder}. The BCF model, which is based on the mechanics of BART but not explicilty designed for confounder selection, is a highly cited method for causal estimation, and is well known for its estimation accuracy in the presence of strong confounding and for its ability to characterize heterogeneous treatment effects. BCF requires an estimate of the propensity score in conjunction with the BART outcome model; we implement BCF with an estimate of the propensity score from the correctly specified exposure model (i.e., the propensity score function assuming the true confounders were known) for each scenario (although note that this specification would typically not be known in practice).  We use the R package \verb|bcf| to run the BCF.  The BAC model \citep{wang2012bayesian, wang2015accounting}) is designed specifically for confounder selection.  It links a model for the exposure and a model for the outcome by introducing a dependence parameter ($\omega$) for the prior odds that a variable included in the exposure model is also included in the outcome model for effect estimation. It relies on parametric specification of the exposure and outcome models. We use the R package \verb|bacr|, specifying all two-way interactions between exposure and every observed covariates in the outcome model and $\omega$ set to $\infty$ (default) which forces predictors in the exposure to be included in the outcome model. The twang model utilizes generalized boosted regression to estimate an exposure model, and uses estimated propensity scores from that model to conduct a weighted comparison of outcomes between exposure groups (i.e., no outcome model is specified). We use the R package \verb|twang| for implementation. The CovSelHigh model \cite{haggstrom2018data} is a probabilistic graphical model where an estimated graph is used to select a set of confounders (and predictors) according to a specified criterion (e.g., the disjunctive cause criterion without instruments) by estimating Markov and Bayesian networks followed by a flexible model for causal effect estimation.  We use the R package \verb|CovSelHigh| to run the model with the network algorithm set to the Max-Min Parents and Children (MMPC) and the effect estimation strategy set to targeted maximum likelihood estimation (TMLE). This model produces three different estimates: (1) an estimate based on a subset of the covariates that cause treatment and dependent with outcome (CovSelHigh$_Q$); (2) an estimate based on a subset of the covariates that cause outcome and dependent with treatment (CovSelHigh$_Z$); and (3) an estimate based on all causes of treatment and/or outcome (CovSelHigh$_{X.TY}$). The BayesPen model first fits the full standard Bayesian regression model and then post-processes the posterior distribution by penalizing models omitting important confounders. This model is fitted using the \verb|R| package \verb|BayesPen| with flat priors on the coefficients of both exposure and outcome models. In each scenario, from the resulting solutions we choose the solution including true confounders with the smallest number of covariates.
\begin{table}[t]
{\renewcommand{\arraystretch}{0.98}
\caption{\footnotesize Simulation results (bias, mean squared error, coverage) under six different scenarios. }
\centering
\resizebox{0.93\columnwidth}{!}{%
\begin{tabular}{|c|c||c|c|c|c|c|c|}
\hline
 \multicolumn{2}{|c||}{}  & Scenario 1 & Scenario 2 & Scenario 3 & Scenario 4 & Scenario 5 & Scenario 6\\
\hline
\multicolumn{2}{|c||}{True Effect} & -2.5656 & -1.3989 & -1.3989 & -2.5656 & -2.5656 & -2.5656\\
\hline
\hline
\multirow{3}{*}{BCF}& Bias & 0.1296& 0.1027& 0.0937&0.2368 & 0.1073 & 0.1371\\
 & MSE& 0.0303 & 0.0201 & 0.0189 & 0.0856& 0.0224 & 0.040\\
 & Coverage& 0.81 &0.77 &0.89 &0.71 &0.63  & 0.75\\
 \hline
\multirow{3}{*}{BAC} & Bias & 0.8398& 0.7709& 0.7900& 0.8564& 0.8469 & 0.3720\\
 & MSE& 0.7474& 0.6471& 0.6627& 0.7684& 0.7409 & 0.1490\\
 & Coverage&0 &0.03 & 0.01& 0& 0 & 0.04\\
 \hline
\multirow{3}{*}{twang} & Bias & 1.8609 & 1.8735& 1.4615& 1.9054 & 1.6284 & 0.6824\\
 & MSE& 3.5686& 3.6202& 2.2388& 3.8067& 2.7529 & 0.5530\\
 & Coverage&0 &0 &0 &0 &0 & 0.33\\
 \hline
\multirow{3}{*}{CovSelHigh$_Q$} & Bias & 0.8713 & 1.0236 & 0.8871 & 1.1143 & 0.7498 & 0.6297\\
 & MSE& 1.0935 & 1.2829& 1.0909& 1.6462& 0.9433 & 0.5294\\
 & Coverage& 0.3 & 0.18 & 0.31& 0.27 & 0.41 &0.31\\
 \hline
\multirow{3}{*}{CovSelHigh$_Z$} & Bias & 1.2320& 1.2857 & 0.8279& 1.6717&  1.3711 & 0.8098\\
 & MSE&1.7733 & 1.8930 & 0.8091 & 3.1997 & 2.3987 & 0.7343\\
 & Coverage&0.11 & 0.05 & 0.25 & 0.07 & 0.1 & 0.0896\\
 \hline
\multirow{3}{*}{CovSelHigh$_{X.TY}$} & Bias & 0.1332 & 0.0829& 0.2322 & 0.3501&  0.1551 & 0.2201\\
 & MSE& 0.0507 & 0.0343 & 0.1812& 0.2528& 0.0742 & 0.1172\\
 & Coverage&0.55 & 0.69 & 0.49& 0.63 & 0.74 & 0.62\\
 \hline
\multirow{3}{*}{BayesPen} & Bias &  0.8184& 0.8135 &0.7135& 0.8454& 0.8344 & 0.3487\\
 & MSE& 0.7037 & 0.6934 &0.5541& 0.7656& 0.7288 & 0.1352\\
 & Coverage& 0.01& 0.01 &0.05& 0.06& 0.01 & 0.15\\
 \hline
\multirow{3}{*}{Ours (Separate)} & Bias & 0.1135 & 0.0946& 0.0649& 0.2223& 0.1034 & 0.1744\\
 & MSE& 0.0232& 0.0158 & 0.0101& 0.0768& 0.0217 & 0.0464\\
 & Coverage& 0.80 & 0.77& 0.84& 0.75& 0.74 & 0.76\\
 \hline
\multirow{3}{*}{Ours (Marginal)} & Bias & \bf 0.0413 & \bf 0.0304& \bf 0.0221&\bf  0.0804& \bf 0.0413 & \bf0.0773\\
 & MSE&\bf 0.0111 & \bf 0.0071& \bf 0.0004 & \bf 0.0315& \bf 0.0127 & \bf0.0205\\
 & Coverage& 0.85& 0.95&0.93 &0.82 & 0.79 & 0.83\\
 \hline
\end{tabular}%
}
}\label{tab:simulation}
\end{table}

\subsection{Simulation Results, $P < N$}
Table 2 summarizes the results from the six simulation settings. In terms of bias and MSE, the proposed marginal model outperforms other methods across all scenarios. Note in particular the marginal model produces biases (and MSEs) three times less than those from the BCF model, a somewhat surprising result since BCF has been shown to perform very well relative to other similar methods for causal effect estimation \citep{hahn2020bayesian}.  However, it is worth noting that the present study does not consider heterogeneous treatment effects (a key strength of BCF), and the default prior specification in BCF is not intended to focus posterior support on a set of necessary covariates that is a small subset of those available for analysis, as is the case here.  Appendix B examines different simulation scenarios designed particularly for the settings where BCF is known to perform well, and illustrates situations where the proposed model performs well compared to the BCF or vice versa.

The poor performance of BAC in this simulation study is not particularly surprising given the fact that the method depends on (generalized-)linear models such that it is hard to capture non-linear terms of the confounders simulated in these scenarios. Similarly for the poor performance of BayesPen in this simulation study, as this method also depends on parametric model structures and thus it does not work properly for complex structures in these simulation scenarios.  The performance of the twang method in this simulation study illustrates the difficulty in including a high dimensional set of variables in a flexible propensity score model, especially when only a small subset are genuinely required for causal effect estimation.  The CovSelHigh model based on all causes of treatment and outcome (CovSelHigh$_{X.TY}$) performs relatively well compared to BAC, twang, BayesPen and its two siblings (CovSelHigh$_Q$, CovSelHigh$_Z$) as illustrated in \cite{haggstrom2018data}. However, the resulting biases are larger than 0.1 for Scenarios 1,3,4, 5, and 6 whereas the proposed marginal model produces biases much smaller than 0.1 across the scenarios. 

\begin{figure}[t]
\centering
 \includegraphics[width=0.85\textwidth]{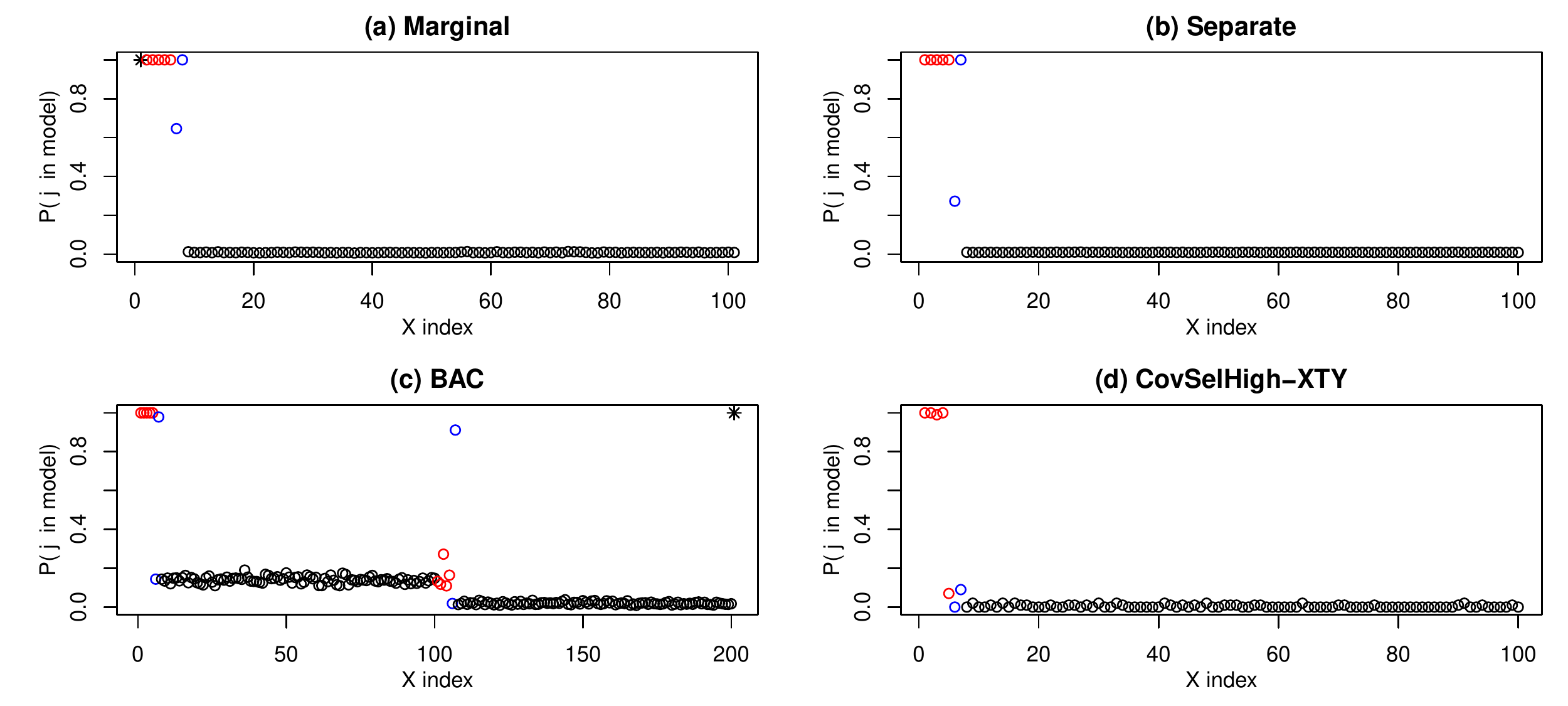}
    \caption{\footnotesize [Scenario 1] The posterior probability of inclusion in the model. Four plots are for the marginal, separate, BAC, and CovSelHigh$_{XTY}$, respectively. Red circles indicate 5 true confounders and blue circles indicate 2 additional predictors in the outcome model. In the marginal model and the BAC model, a black star indicates the exposure variable. In the BAC model, the last 100 points are for interactions between the exposure variable and each confounder.}   \label{fig:simulation1}
\end{figure}

\begin{figure}[t]
\centering
 \includegraphics[width=0.85\columnwidth]{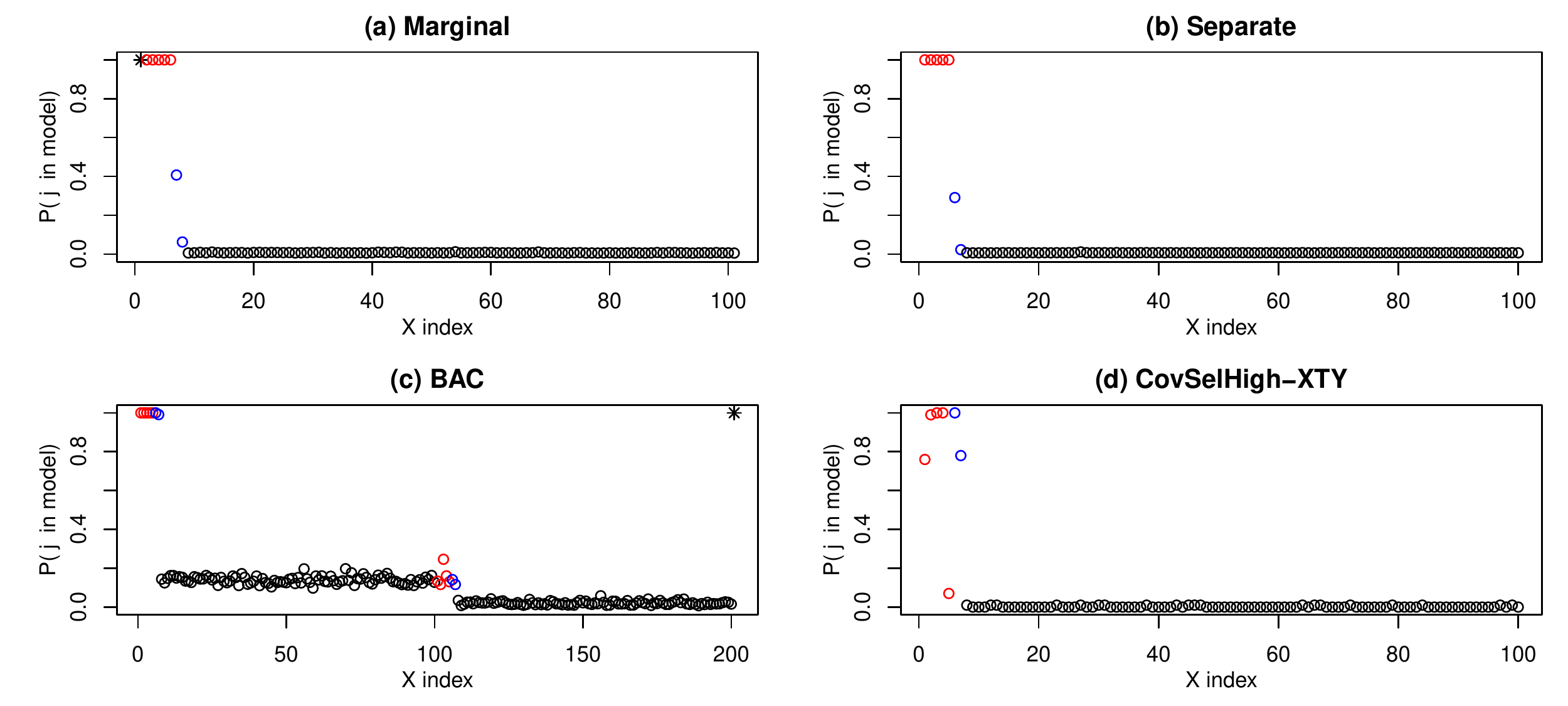}
    \caption{\footnotesize [Scenario 3] The posterior probability of inclusion in the model. Four plots are for the marginal, separate, BAC, and CovSelHigh$_{XTY}$, respectively. Red circles indicate 5 true confounders and blue circles indicate 2 instrumental variables in the exposure model. In the marginal model and the BAC model, a black star indicates the exposure variable. In the BAC model, the last 100 points are for interactions between the exposure variable and each confounder.}   \label{fig:simulation2}
\end{figure}

\begin{figure}[t]
\centering
 \includegraphics[width=0.85\columnwidth]{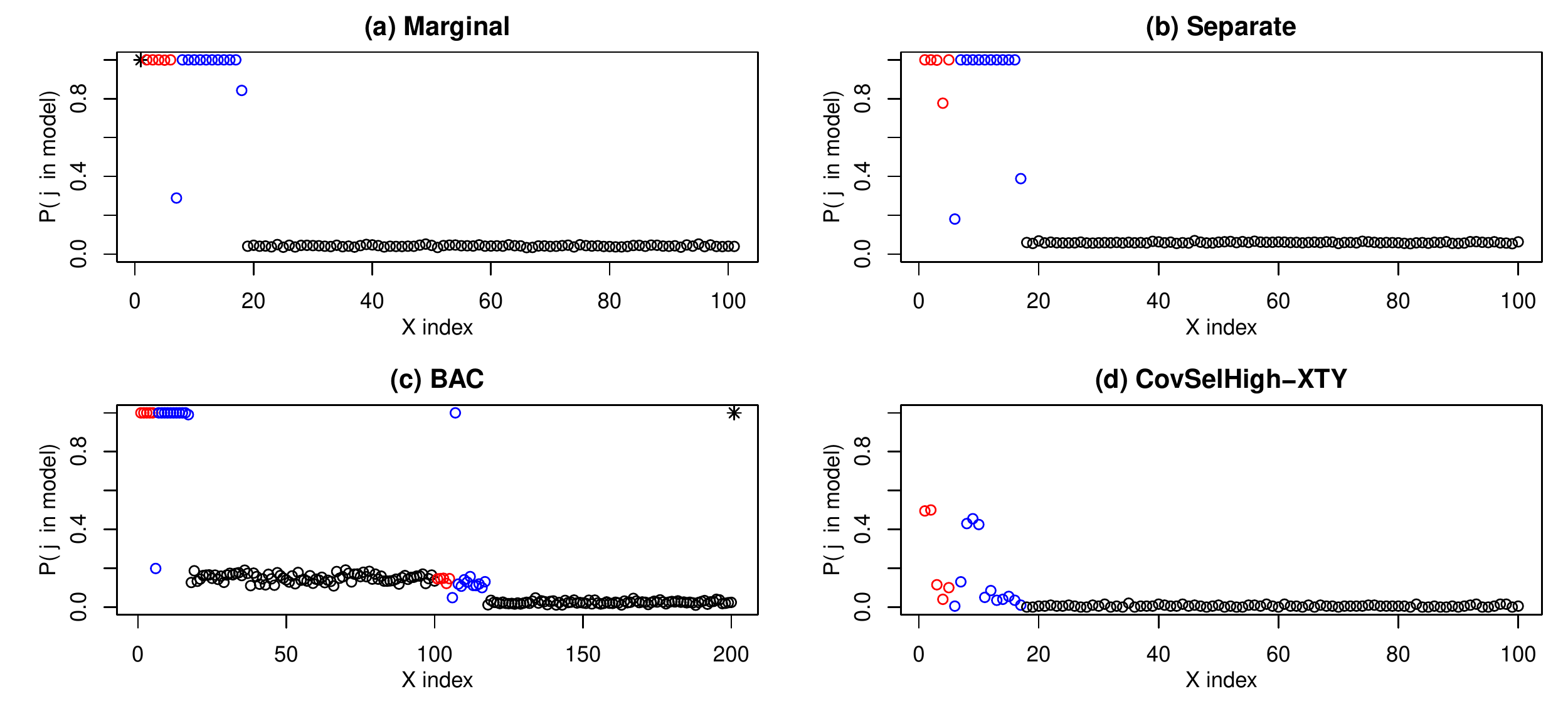}
    \caption{\footnotesize [Scenario 6] The posterior probability of inclusion in the model. Four plots are for the marginal, separate, BAC, and CovSelHigh$_{XTY}$, respectively. Red circles indicate 5 true confounders and blue circles indicate 12 additional predictors in the outcome model. In the marginal model and the BAC model, a black star indicates the exposure variable. In the BAC model, the last 100 points are for interactions between the exposure variable and each confounder.}   \label{fig:simulation3}
\end{figure}

To illustrate how the various methods prioritize different variables, Figures \ref{fig:simulation1}-\ref{fig:simulation3} show the posterior probabilities of inclusion for all covariates in the proposed, BAC, and CovSelHigh methods across simulation scenarios 1, 3, and 6. The posterior probability of inclusion is defined as $\frac{1}{250}\sum_{m=1}^{250} \bar{P}_m(j)$ for each $j$ potential confounder where $\bar{P}_m(j)$ denotes the posterior probability of inclusion of the $j$-th confounder in the model for the $m$-th simulation replicate. Across all three of these scenarios, the proposed methods (the top two panels in each figure) select true confounders (red points) with posterior probability 1, with the single exception in Scenario 6 where the confounder with the weakest association with the outcome ($X_4$) has average posterior inclusion probability around 0.8. The BAC method similarly exhibits very high posterior probability of inclusion for the main effects of the true confounders, but struggles to select interaction terms involving these confounders, likely a consequence of its reliance on the need to specify a functional form of the interaction {\it a priori} that, in this case, does not reflect the data generation. The ability of the CovSelHigh algorithms to include the true confounders is more variable across these scenarios, indicating substantial posterior probability of missing confounders in Scenario 6 when the associations are smaller than in the other scenarios.  Figure \ref{fig:simulation1} shows how the proposed methods are more likely than the comparison methods to include predictors of $Y$ unassociated with $A$. Figure \ref{fig:simulation2} illustrates how the proposed methods satisfy the disjunctive cause without instruments criterion. The proposed methods give lower than a 50\% posterior probability of inclusion for each of two instruments \citep{barbieri2004optimal}, but assign posterior probability 1 to all of true confounders. Specifically, our proposed models assigns posterior probability of 1 to the first term in Eq.(\ref{decomp}) (i.e., $\mathcal{I}_\cap$), and the BAC also assigns posterior probability of 1 to the models in $\mathcal{I}_\cap$. If we further examine the nature of the models constituting $\mathcal{I}_\cap$, our proposed separate model (and marginal model) assigns 0.70 (and 0.60) to the models in $\mathcal{I}_\star$ (i.e., the first term in Eq. (\ref{decomp1})) while the BAC assigns 0 posterior probability to the models in that class, meaning that in the presence of an unmeasured confounder and the threat of Z-bias, our methods can weight more the models that produce unbiased effect estimates. Figure \ref{fig:simulation3} shows how the methods perform when associations are comparatively lower than the other scenarios.  Here, we see that the proposed method continues to assign very high posterior inclusion probability to true confounders and most predictors of $Y$, that BAC continues to exclude interaction terms involving confounders and predictors, and that the performance of both CovSelHigh algorithms deteriorates.

\subsection{$P > N$ scenario}
We perform an additional simulation study to evaluate the ability of the proposed model when the number of potential confounders $(P=100)$ is strictly larger than the sample size $(n=60)$. To test the confounder selection and effect estimation performance in this setting we use the same setting in Scenario 2 except the sample size $n$ set to $60$. Due to the $P > N$ high-dimensional setting, some approaches cannot initiate the corresponding R functions (i.e., BAC and BayesPen). For the BCF model, we only consider the first 50 covariates ($P^\star=50$) for the prognostic function $\mu(x)$. Table \ref{tab:simulation1} shows the results. Both of our proposed models outperform in terms of biases and MSEs.
\begin{table}
{\renewcommand{\arraystretch}{1}
\caption{\footnotesize Simulation results (bias, mean squared error, coverage) when $P > n$. $\dagger$ BCF only uses the first 50 covariates $x_1 - x_{50}$ for the prognostic function $\mu(x)$.}\label{tab:simulation1}
\begin{center}
\resizebox{.95\textwidth}{!}{%
\begin{tabular}{|l|c|c|c|c|c|c|c|}
\hline
   & BCF$^{\dagger}$ & twang & CovSelHigh$_Q$ & CovSelHigh$_Z$ & CovSelHigh$_{X.TY}$ &  Ours (Separate) & Ours (Marginal) \\
 \hline
  Bias & 1.2109 & 2.1100  &1.9976 & 2.2527 & 1.8000  &  1.3304 & \bf 0.9146\\
   MSE & 1.5701& 5.0995  & 4.6098 & 5.6062 & 3.7286   & 2.0982 & \bf 1.4597\\
 Coverage & 0.34 & 0.24 & 0.15 & 0.05 & 0.06  & 0.30 & \bf 0.57 \\
\hline
\end{tabular}%
}
\end{center}
}
\end{table}

\section{Estimating Causal Effects of Power Plant Emissions on Ambient Particulate Pollution}    
We estimate the causal effect of low HyADS coal-emissions exposure ($A$) on annual ambient PM$_{2.5}$ concentrations ($Y$) in 2013. Recall that we have 104 potential confounders available in the final data set (after excluding variables that are strongly correlated with the remaining potential confounders), which includes variables specific to each location as well as regional information at various directions located 100km from the ZIP code. Table \ref{tab:result} includes posterior mean estimates (with 95\% posterior intervals) for the causal effect from five different methods: our marginal and separate models, BCF from Hahn et al. (2020), BAC from Wang et al. (2015), and twang from Ridegeway et al. (2008). We exclude CovSelHigh from H\"{a}ggstr\"{o}m (2018) and BayesPen from Wilson and Reich (2014) from this analysis because the former evaluates all possible network connections among variables and easily exceeds the memory limit with a larger set of confounders, and the latter requires computation of $(X^\top X)^{-1}$ for the model with all available covariates, which is difficult to invert with multicollinearity (e.g., multiple temperature measures for each locations). We use all available covariates as potential confounders in each of five competing models to allow uncertainty around confounder selection. For our marginal and separate models, the number of distinct trees is set to 200 for each model and for the (hyper-)parameters, we use the recommended settings in \cite{kapelner2013bartmachine}. We run 2 MCMC chains each with 100,000 iterations and discard the first half as burn-in. To check MCMC convergence, we examine the Gelman-Rubin diagnostics for multiple chains and indicates that no critical convergence issues are detected (1.2 for the marginal model and 1.1 for the separate model). We also use graphical posterior predictive checks in Figure S2 in the supplementary material to determine the model fit by displaying observed data $y$ along with 7 replicated datasets $y^\text{pred}$ from the posterior predictive distributions of the separate and marginal models, which suggests that our proposed models fit the observed data well.

Our approaches estimate that the low coal-emissions exposure causes reduction in ambient PM$_{2.5}$ concentrations by -0.07 (-0.13, -0.04) mg/m$^3$ and -0.64 (-0.67, -0.61) mg/m$^3$ for the marginal model and the separate model, respectively. This indicates that ambient PM$_{2.5}$ concentrations are affected by the HyADS coal-emissions exposure on average, especially when allowing the model for PM$_{2.5}$ to differ in the low and high HyADS exposure areas. The BCF model produces the posterior mean estimate that lies between those from our marginal model and separate model, albeit with a wider posterior interval than that from the marginal model, suggesting the potential benefit of the type of confounder selection pursued here. These results can be compared with the previous analysis in \citep{kim2020health}, which considers an earlier year (2005) when coal emissions pollution was much higher and a larger contributor to PM$_{2.5}$ pollution than in 2013 and also considered a smaller set of confounders, estimating that a low coal-emissions exposure causes reduction in ambient PM$_{2.5}$ concentrations by -1.75 (-1.80, -1.71) mg/m$^3$.  The twang model also suggests that reduction in ambient PM$_{2.5}$ concentrations is caused by lowering coal-emissions exposure. However, its estimate (and interval) is significantly different from those produced by our proposed methods or BCF, likely a result of the erratic behavior of inverse weighting estimators with many unnecessary variables included in the propensity score model. The BAC model estimates that the low coal-emissions exposure increases ambient PM$_{2.5}$ concentrations, which is likely due in this case to the algorithm frequently involving MCMC iterations with an exposure model producing complete separation of exposure levels.

\subsection{Inference for variable importance}
Figures S3-S4 in the supplementary material depict the posterior probability of inclusion for each covariate in the proposed models (the marginal and separate models). Considering variables with at least a 100\% posterior probability of appearing in the trees as variables that satisfy the disjunctive cause without instruments criterion, the method identifies 51 variables (50 covariates + 1 exposure variable) from the marginal model and 85 covariates from the separate model. The marginal model uses a smaller set of confounders and more restrictive response surface, which results in a narrower posterior interval. 
Even though the two methods have a similar pattern in terms of which variables are included, some of variables play different roles in the two models. Table S3 (and Figures S3 and S4) in the supp. material shows the variables included in each model by category, with the similar number of local weather variables included in each model but a much larger number of regional variables included in the separate model.

Examining the posterior inclusion probabilities provides a sense of the importance of both local and regional meteorological factors, with both of our proposed models including many weather variables at various regional directions as potential confounding variables.  The separate model includes almost all neighboring weather variables (66 regional variables with posterior inclusion probability $= 100\%$) as well as 11 (6 winter and 5 summer) local weather variables. The marginal model prioritizes a smaller set of regional weather variable, with 36 regional weather variables having posterior inclusion probability $= 100\%$.  From the results, we can deduce that the importance of the regional weather variable in this causal estimation is quite large relative to the local meteorological conditions. 



The BAC model identifies 106 terms with posterior inclusion probability $= 100\%$ where 36 of them are interaction terms with the exposure variable (see Figure S5 in supplementary material).  
The number of local weather variables used as a confounder is 19 (including interactions with the exposure), while the number of regional weather variables used in the model is 77 (including interactions with the exposure).

\begin{table}[t]
{\renewcommand{\arraystretch}{1.1}
\caption{\footnotesize Estimates of the causal effect of low HyADS coal-emissions exposure on annual ambient PM$_{2.5}$ concentrations from five different methods for years 2013 and 2014.}\label{tab:result}
\scriptsize
\begin{center}
\begin{tabular}{c|c|ccccc}
\hline
Year& & Ours (Marginal) & Ours (Separate) & BCF & BAC & twang\\
\hline
\multirow{2}{*}{2013}& Posterior Mean & -0.07 & -0.64 & -0.28 & 0.59 & -1.12\\
&Posterior 95\% C.I. & (-0.13, -0.04) & (-0.67, -0.61) & (-0.39, -0.19) & (0.47, 0.71) & (-1.17, -1.07)\\
\hline 
\multirow{2}{*}{2014}& Posterior Mean & -0.08 & -0.43 & 0.04 & -0.28 & -1.26\\
&Posterior 95\% C.I. & (-0.12, -0.07) & (-0.48, -0.38) & (-0.08, 0.16) & (-0.39, -0.19) & (-1.31, -1.20)\\
\hline 
\end{tabular}

\end{center}
}
\end{table}

\begin{figure}[t]
\centering
 \includegraphics[width=0.55\textwidth]{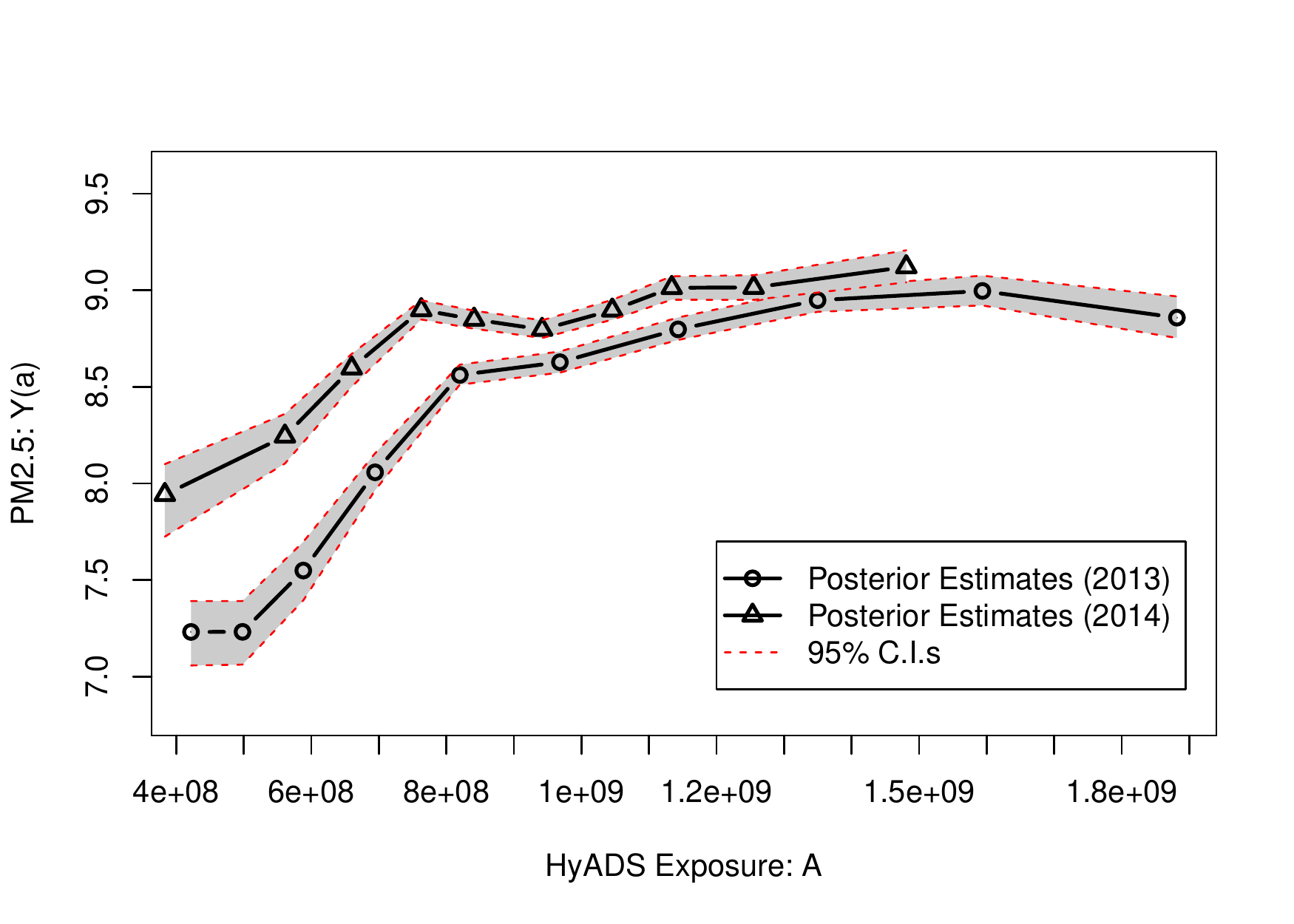}
    \caption{\footnotesize The exposure-response surface plot for years 2013 and 2014, which is obtained by marginalizing the posterior samples of the (continuous) marginal model over all confounders. Note that, for the binary exposure case, we use the cutoff exposure value of 980,249,908.}     \label{fig:ersurface}
\end{figure}

\subsection{Additional analyses}
We conduct two additional analyses: (a) an analysis identical to that in Section 7.1, but for year 2014 (using the same cutoff value to dichotomize the exposure) to check whether the findings from the previous analysis were consistent with the analysis of results from an adjacent year with similar emissions conditions, and (b) analyses for year 2013 and 2014 with the exposure, $A$, regarded as continuous where, as illustrated in Section 3.3, we replace the binary exposure model (Eq. 2) with a BART model for continuous exposure variable.

Figure S6 in the supplementary material illustrates the distribution of HyADS coal-emissions exposure levels and the treated (blue) and control (red) zip code locations in 2014 dichotomized based on the mean of HyADS coal emissions exposure levels in 2013. In 2014, more zip locations are assigned to the low HyADS exposure group ($A=1$; 7335 locations) and fewer zip code locations are allocated to the high HyADS exposure group ($A=0$; 5608 locations) suggesting that the overall HyADS exposure level decreases in 2014. Table \ref{tab:result} illustrates the estimates for year 2014. The marginal and separate model estimate that the low HyADS emissions exposure causes reduction in ambient PM$_{2.5}$ concentrations by -0.08 (-0.12, -0.07) mg/m$^3$ and -0.43 (-0.48, -0.38) mg/m$^3$, respectively, representing close agreement with the analysis from 2013 in Section 7.1. For the 2014 analysis, the BCF and BAC models produce estimates that differ in important ways from their 2013 counterparts. BCF produces a posterior mean estimate of 0.04 with a wider posterior interval that includes 0. BAC estimates that the low coal-emissions exposure causes reduction in ambient PM$_{2.5}$ concentrations by -0.28 (-0.39, -0.19) mg/m$^3$, but still with a wider posterior interval. The estimate from the twang model in 2014 is more similar to its corresponding 2013 estimate, but remains significantly different from the estimates from the other approaches.

When considering $A$ as a continuous exposure variable, we can only consider the marginal model since the separate model is built on two outcome models for $A=0, 1$. Here, we consider the target causal estimand $\Delta(a, a^\prime)$ for different exposure levels $a \neq a^\prime$. For 2013, when $a = \text{25th percentiles}=748882856$ and $a^\prime = \text{75th percentiles}=1189376706$, it is estimated that the low coal-emissions exposure causes reduction in ambient PM$_{2.5}$ concentrations by -0.59 (-0.70, -0.49) mg/m$^3$. When $a = \text{50th percentile}=927708796$ and $a^\prime = \text{75th percentiles}=1189376706$, it is estimated that the low coal-emissions exposure causes reduction in ambient PM$_{2.5}$ concentrations by -0.23 (-0.29, -0.14) mg/m$^3$. For 2014, when $a = \text{25th percentiles}=688843769$ and $a^\prime = \text{75th percentiles}=1141206250$, it is estimated that the low coal-emissions exposure causes reduction in ambient PM$_{2.5}$ concentrations by -0.26 (-0.35, -0.14) mg/m$^3$. When $a = \text{50th percentiles}=914264627$ and $a^\prime = \text{75th percentiles}=1141206250$, it is estimated that the low coal-emissions exposure causes reduction in ambient PM$_{2.5}$ concentrations by -0.15 (-0.21, -0.07) mg/m$^3$. Figure \ref{fig:ersurface} depicts the exposure-response function $E[Y(a)]$ of varying $A=a$ values for the marginal models fit to data from both 2013 and 2014, which are the Partial Dependence Plot with respect to exposure $A$. This indicates that Ambient PM$_{2.5}$ is most sensitive to changes in HyADS emissions exposure at levels below an apparent threshold of approximately 8e+08, above which changes in emissions have relatively minor effects on ambient PM2.5.


\section{Discussion}
In this article, we present a Bayesian additive regression trees model to estimate causal effects from observational data when a number of potential confounders is large relative to the sample size, indicating potential benefits of concerted efforts to prioritize which of a large set of potential confounders should be included in the analysis. The proposed method allows estimation of causal effects with such prioritization of the relevant variables, while accounting for uncertainty in confounder selection. Since the proposed method does not depend on a parametric model assumption, it can handle any data model having complex non-linear dose-response functions. The simulation results show that the proposed method outperforms many existing methods in terms of lower biases and MSEs, and consistency of results in two analyses of power plant pollution in consecutive years points towards successful adjustment for confounding relative to other similar methods. 

Our methods estimate that low HyADS coal-emissions exposure reduces ambient PM$_{2.5}$ concentrations in the Eastern United States in 2013 and 2014. When compared to other competing methods, our two methods' estimates are consistent with each other, with narrower posterior intervals, indicating the benefit of concerted efforts towards confounder selection. The number of confounders identified with a posterior inclusion probability of 100\% is lower in our proposed methods (particularly in the marginal model), reducing variability in the estimates.

The proposed method of confounder selection is closely linked to the confounder selection principle described the ``disjunctive cause criterion without instruments," that is, the method is able to prioritize confounders {\it and} variables that are predictive of outcomes, while de-prioritizing inclusion of instruments. Therefore, we view subsets of variables prioritized by the proposed methodology as optimal target sets of confounders that produce unbiased estimates of casual effects.

There are several avenues for future work to refine the methodology proposed here. First, the methods proposed here use the default settings for the component BART model (hyper-) parameters. Even though we find no evidence that causal estimates in the cases considered here  heavily depend on the parameter settings, we could alternatively use a cross-validation to tune the (hyper-)parameters for more precise estimates. The principles outlined here might also extend to settings where there are more causal variables of interest, for example, in the mediation analysis context.  The shared prior distribution across BART models could be extended to a third model for a mediating variable to  include information about which variables are confounders in outcome-mediator and/or mediator-exposure relationships. It would be also interesting to develop a strategy to give a constraint on the selection probabilities when we want to force some covariates to be in the model based on some prior knowledge.

There are also potential extensions to incorporate the confounder selection capabilities explored here into the BCF framework. In principle, a common splitting prior could be incorporated for confounder selection in both the propensity score and BCF outcome model, but a key challenge is simultaneously estimating the propensity score function and the prognostic function since the latter is a function of the former. Possible solutions might come from literature on so-called ``Bayesian propensity scores'' 
\citep{zigler2016central}, where various approaches to ``cut the feedback'' \citep{mccandless2010cutting} have been documented, as well as methodologies based on the Bayesian bootstrap \citep{stephens2022causal}.


\bigskip
\begin{center}
{\large\bf SUPPLEMENTARY MATERIAL}
\end{center}

\begin{description}

\item[Web-based Supplementary Material:] Web appendices and Figures (which further describe posterior computation, simulation set-up, model checking, and results) referenced in Sections 4.1, 5, 6 and 7. (.pdf file)

\item[R-code:] R-code to perform the proposed methods under six simulation scenarios.

\end{description}

\bibliographystyle{agsm}

\begingroup
\setstretch{1}
\bibliography{BART_CS}
\endgroup
\end{document}


\def\spacingset#1{\renewcommand{\baselinestretch}%
{#1}\small\normalsize} \spacingset{1}


\if1\blind
{
  \title{Web-based Supplementary Materials for ``Bayesian Nonparametric Adjustment of Confounding'' by }
 \author{\small Chanmin Kim\\
   \small Department of Statistics, SungKyunKwan University, Seoul, Korea\\
    \small and \\
    \small Mauricio Tec \\
    \small Department of Statistics and Data Science, The University of Texas, Austin, TX 78712 \\
    \small and \\
    \small Corwin M. Zigler \\
    \small Department of Statistics and Data Science, The University of Texas, Austin, TX 78712 }
    \date{}
  \maketitle
} \fi

\if0\blind
{
  \bigskip
  \bigskip
  \bigskip
  \begin{center}
    {\LARGE\bf Bayesian Nonparametric Adjustment of Confounding}
\end{center}
  \medskip
} \fi

\bigskip

\spacingset{1.9} 

\section*{Appendix A: Acceptance Ratios}
We use the following acceptance ratios \citep{kapelner2013bartmachine} for three tree alteration proposals with $P_\text{grow} = 0.28, P_\text{prune} = 0.28,P_\text{change} = 0.44$.
\begin{itemize}
\item Grow proposal:
\begin{eqnarray*}
r &=& \frac{P_{\text{prune}}}{P_{\text{grow}}}\frac{b p_\eta n_{p,\eta}}{w} \sqrt{\frac{\sigma^2(\sigma^2+n_\eta\sigma_{\mu}^2)}{(\sigma^2+n_{\eta_L}\sigma_\mu^2)(\sigma^2+n_{\eta_R}\sigma_\mu^2)}} \\
& & \times \exp\left(\frac{\sigma_\mu^2}{2\sigma^2}\left(\frac{(\sum_{i=1}^{n_{\eta_L}}R_{\eta_L,i})^2}{\sigma^2+n_{\eta_L} \sigma^2_\mu}+\frac{(\sum_{i=1}^{n_{\eta_R}}R_{\eta_R,i})^2}{\sigma^2+n_{\eta_R} \sigma^2_\mu}-\frac{(\sum_{i=1}^{n_{\eta}}R_{\eta,i})^2}{\sigma^2+n_{\eta} \sigma^2_\mu}\right)\right)\\
& & \times \frac{\beta_1 (1-\frac{\beta_1}{(2+d_\eta)^{\beta_2}})^2}{((1+d_\eta)^{\beta_2}-\beta_1)p_\eta n_{p,\eta}}
\end{eqnarray*}
where $b$ is the number of terminal nodes in current tree $\mathcal{T}_j$, $p_\eta$ denotes the number of predictors available to split on, $w^\star$ denotes the number of singly internal nodes in a new tree, $n_{p,\eta}$ is the number of unique values left for the selected predictor. $\sigma^2_\mu$ is the prior variance for the mean parameter in each terminal node, $R_\eta$ and $n_\eta$ denotes the data in the $\eta$th terminal node and the number of observations in that node, respectively. In the grow proposal, the $\eta$th terminal node generates two children, which are denoted by $\eta_l$ and $\eta_R$. 
Here, $\beta_1$ and $\beta_2$ are parameters for the probability of splitting on a given node, $\beta_1 / (1+d_\eta)^{\beta_2}$ where $d_\eta$ is the depth of the node. 
\item Prune proposal:
\begin{eqnarray*}
r &=& \frac{P_{\text{grow}}}{P_{\text{prune}}}\frac{w}{(b-1)p_{\eta^\star}n_{p,\eta^\star}}\sqrt{\frac{(\sigma^2+n_{\eta^\star_L}\sigma_\mu^2)(\sigma^2+n_{\eta^\star_R}\sigma_\mu^2)}{\sigma^2(\sigma^2+n_\eta^\star\sigma_{\mu}^2)}}\frac{\left((1+d_\eta)^{\beta_2}-\beta_1\right)p_{\eta^\star}n_{p,\eta^\star}}{\beta_1\left(1-\frac{\beta_1}{(2+d_\eta)^{\beta_2}}\right)}\\
& &\times \exp\left(\frac{\sigma_\mu^2}{2\sigma^2}\left(-\frac{(\sum_{i=1}^{n_{\eta^\star_L}}R_{\eta^\star_L,i})^2}{\sigma^2+n_{\eta^\star_L} \sigma^2_\mu}-\frac{(\sum_{i=1}^{n_{\eta^\star_R}}R_{\eta^\star_R,i})^2}{\sigma^2+n_{\eta^\star_R} \sigma^2_\mu}+\frac{(\sum_{i=1}^{n_{\eta^\star}}R_{\eta^\star,i})^2}{\sigma^2+n_{\eta^\star} \sigma^2_\mu}\right)\right)
\end{eqnarray*}
where $w$ denotes the number of singly internal nodes in the current tree, and $\eta^\star$ denotes the node becoming a new terminal node in the proposed tree.
\item Change proposal:
\begin{eqnarray*}
r &=&  \sqrt{\frac{\left(\frac{\sigma^2}{\sigma^2_\mu}+n_1\right)\left(\frac{\sigma^2}{\sigma^2_\mu}+n_2\right)}{\left(\frac{\sigma^2}{\sigma^2_\mu}+n_1^\star\right)\left(\frac{\sigma^2}{\sigma^2_\mu}+n_2^\star\right)}} \\
& & \times \exp\left(\frac{1}{2\sigma^2} \left(\frac{(\sum_{i=1}^{n_1^\star} R_{1^\star, i})^2}{n_1^\star+ \frac{\sigma^2}{\sigma^2_\mu}}+\frac{(\sum_{i=1}^{n_2^\star} R_{2^\star, i})^2}{n_2^\star+ \frac{\sigma^2}{\sigma^2_\mu}}-\frac{(\sum_{i=1}^{n_1} R_{1, i})^2}{n_1+ \frac{\sigma^2}{\sigma^2_\mu}}-\frac{(\sum_{i=1}^{n_2} R_{2, i})^2}{n_2+ \frac{\sigma^2}{\sigma^2_\mu}}\right)\right),
\end{eqnarray*}
where $R_1^{\star}$ and $R_1$ are the residuals of the first child node in the proposed and the current trees, respectively, and $n_1^\star$ and $n_1$ are their corresponding number of observations. Similarly, $R_2^\star$ (and $n_2^\star$) and $R_2$ (and $n_2$) are for the second child node in the proposed and current trees.
\end{itemize}

\subsection*{Appendix B: Simulation Setup}
In six different scenarios, 100 potential confounders ($X_1-X_{100}$) are independently generated from $N(0,1)$ where only 5 of them ($X_1-X_5$) are true confounders:
\begin{itemize}
\item Scenario 1: $A$ model contains only true confounders ($X_1 - X_5$); $Y$ model contains true confounders ($X_1 - X_5$) and predictors ($X_6-X_7$).
{\footnotesize
\begin{eqnarray*}
P(A_i=1) &=& \Phi(0.5+0.5 h_1(X_{i,1}) + 0.5 h_2(X_{i,2}) - 0.5 |X_{i,3}-1| +1.5 X_{i,4} X_{i,5})\\
Y_i & \sim & N(\mu(\boldsymbol{X}_i) ,\, 0.3^2) \\
\mu(\boldsymbol{X}_i) & = & h_1(X_{i,1}) + 1. 5 h_2(X_{i,2}) - A_i + 2 |X_{i,3}+1| + 2 X_{i,4} + \exp(0.5 X_{i,5}) - 0.5 A_i |X_{i,6}| - A_i |X_{i,7}+1|
\end{eqnarray*}
}
where $h_1(x) = (-1)^{I(x<0)}$ and $h_2(x) = (-1)^{I(x\geq0)}$.
\item Scenario 2: $A$ model contains only true confounders ($X_1 - X_5$); $Y$ model contains true confounders ($X_1 - X_5$).
{\footnotesize
\begin{eqnarray*}
P(A_i=1) &=& \Phi(0.5+0.5 h_1(X_{i,1}) + 0.5 h_2(X_{i,2}) - 0.5 |X_{i,3}-1| +1.5 X_{i,4} X_{i,5})\\
Y_i & \sim & N( h_1(X_{i,1}) + 1. 5 h_2(X_{i,2}) - A_i + 2 |X_{i,3}+1| + 2 X_{i,4} + \exp(0.5 X_{i,5}) - 0.5 A_i |X_{i,5}|,\, 0.3^2)
\end{eqnarray*}
}
\item Scenario 3: $A$ model contains true confounders ($X_1 - X_5$) and `instrumental variables' ($X_6 - X_7$); $Y$ model contains true confounders ($X_1 - X_5$).
{\footnotesize
\begin{eqnarray*}
P(A_i=1) &=& \Phi(0.5+0.5 h_1(X_{i,1}) + 0.5 h_2(X_{i,2}) - 0.5 |X_{i,3}-1| +1.5 X_{i,4} X_{i,5}+ 1.5 X_{i,6} - X_{i,7})\\
Y_i & \sim & N( h_1(X_{i,1}) + 1. 5 h_2(X_{i,2}) - A_i + 2 |X_{i,3}+1| + 2 X_{i,4} + \exp(0.5 X_{i,5}) - 0.5 A_i |X_{i,5}|,\, 0.3^2)
\end{eqnarray*}
}
\item Scenario 4: $A$ model contains only true confounders ($X_1 - X_5$); $Y$ model contains true confounders ($X_1 - X_5$) and predictors ($X_6-X_{17}$).
{\footnotesize
\begin{eqnarray*}
P(A_i=1) &=& \Phi(0.5+0.5 h_1(X_{i,1}) + 0.5 h_2(X_{i,2}) - 0.5 |X_{i,3}-1| +1.5 X_{i,4} X_{i,5})\\
Y_i & \sim & N( \mu(\boldsymbol{x}_i),\, 0.3^2)\\
\mu(\boldsymbol{x}_i) &=&  h_1(X_{i,1}) + 1.5 h_2(X_{i,2}) - A_i + 2 |X_{i,3}+1| + 2 X_{i,4} + \exp(0.5 X_{i,5}) - 0.5 A_i |X_{i,6}| \\ 
& & - A_i |X_{i,7}+1| + X_{i,8}+X_{i,9}+X_{i,10}+0.5X_{i,11}+0.5X_{i,12}+0.5X_{i,13}-0.5X_{i,14} \\
& & -0.5X_{i,15}-0.5X_{i,16}-\exp(0.2X_{i,17})
\end{eqnarray*}
}
\item Scenario 5: $A$ model contains only true confounders ($X_1 - X_5$); $Y$ model contains true confounders ($X_1 - X_5$) and predictors ($X_6-X_{17}$). The outcome and exposure models are the same in Scenario 4, but the sample size is $N=500$.
\item Scenario 6: the same settings in Scenario 4 except the coefficients of the true confounders in the $Y$ model are relatively smaller.
{\footnotesize
\begin{eqnarray*}
Y_i & \sim & N( \mu(\boldsymbol{x}_i),\, 0.3^2)\\
\mu(\boldsymbol{x}_i) &=&  0.5 h_1(X_{i,1}) + 0.5 h_2(X_{i,2}) - A_i + 0.5 |X_{i,3}+1| + 0.3 X_{i,4} + \exp(0.5 X_{i,5}) - 0.5 A_i |X_{i,6}|\\
&& - A_i |X_{i,7}+1| + X_{i,8}+X_{i,9}+X_{i,10}+0.5X_{i,11}+0.5X_{i,12}+0.5X_{i,13}-0.5X_{i,14}\\
& &-0.5X_{i,15}-0.5X_{i,16}-\exp(0.2X_{i,17})
\end{eqnarray*}
}
\end{itemize}

\begin{table}[h]
{\renewcommand{\arraystretch}{0.8}
\caption{\footnotesize Six scenarios in the simulation study. $\boldsymbol{S}$ indicates Scenario and () indicates extra covariates used in $Y$ or $A$ models. $\star$ is for the scenario with $n=500$. }\label{tab:scenario}
\begin{center}
{\footnotesize
\begin{tabular}{|c|c|c|c|}
\hline
& &  \multicolumn{2}{c|}{$Y$ Model}\\
\cline{3-4}
& & $X_1 - X_5$ & $X_1-X_5$ and more\\
\hline
\multirow{4}{*}{$A$ Model}& \multirow{3}{*}{$X_1-X_5$} & & $\boldsymbol{S1}$ ($X_6-X_7$)\\
                                   &   & $\boldsymbol{S2}$ & $\boldsymbol{S4}$ ($X_6-X_{17}$)\\
                                   &   &  & $\boldsymbol{S5}$$^\star$ ($X_6-X_{17}$)\\
                                                                      &   &  & $\boldsymbol{S6}$ ($X_6-X_{17}$)\\
\cline{2-4}
& $X_1-X_5$ and more & $\boldsymbol{S3}$ ($X_6-X_7$) &\\
\hline
\end{tabular}}
\end{center}
} 
\end{table}

\subsection*{Appendix C: BCF vs BART}
In this section, we further investigate the nature of the out-performance of our proposed model compared to the BCF model which is known to perform better than the BART model in the estimation of causal effects. We run an extra simulation study based on the scenario from \cite{hahn2020bayesian} where confounding is driven by so-called ``targeted selection'' - individuals selecting into treatment based on expected outcomes under no treatment. In this scenario with ($n=250$), $X_p$'s are independently generated from $N(0,1)$ for $p=1,2, \cdots, P$ where $X_1$ and $X_2$ are true confounders. The propensity score $\pi = \Phi(\mu(\boldsymbol{x}))$ is a function of the prognostic function $\mu(\boldsymbol{x}) = -I(x_1 > x_2) + I(x_1 < x_2)$. The treatment is generated from $A \sim Bern(\pi)$ and the outcome is generated based on
\[y = \mu(\boldsymbol{x})+\tau(\boldsymbol{x})a + \epsilon, \,\, \epsilon \sim N(0, \sigma^2)\]
where $\tau(\boldsymbol{x})$ is set to $-1$, implying a homogeneous effect. From the general BCF model structure, 
\[E(Y_i|x_i, A_i=a_i) = \mu(\boldsymbol{x}_i, \hat{\pi}(\boldsymbol{x}_i))+\tau(\boldsymbol{x_i})a_i,\]
we consider five different specifications:
\begin{itemize}
\item BCF : all $P$ potential confounders in $\mu, \tau$ and $\pi$ functions.
\item BCF 1 : all $P$ potential confounders in $\mu, \pi$ functions. $\tau$ is constant (homogeneous effect).
\item BCF 2 : all $P$ potential confounders in $\mu$. $\pi$ function is correctly specified with true confounders $x_1$ and $x_2$. $\tau$ is constant
\item BCF 3 : all $P$ potential confounders in $\pi$. $\mu$ function is correctly specified with true confounders $x_1$ and $x_2$. $\tau$ is constant
\item BCF 4 : $\pi$ and $\mu$ functions are correctly specified with true confounders $x_1$ and $x_2$. $\tau$ is constant
\end{itemize}

Table \ref{bcf_bart} summarizes the results under two settings: $P=50$ and $P=100$. When true confounders are provided in either the prognostic function or the propensity score function (BCF 2 - BCF 4), the BCF model outperforms in terms of both bias and MSE. However, our focus is on selection of true confounders from a set of many potential confounders and if that is the case, our method outperforms the BCF models (BCF and BCF 1).

\begin{table}
\caption{Simulation results (bias, mean squared error) under two different settings: the number of potential confounders $P=50$ and  the number of potential confounders $P=100$.}\label{bcf_bart}
\centering
\begin{tabular}{|c|c||c|c||c|c|c|c|}
\hline
 \multicolumn{2}{|c||}{}  & Ours (Marginal)  & BCF & BCF 1& BCF 2 & BCF 3 & BCF 4\\
\hline
\hline
\multirow{2}{*}{P=50}& Bias & 0.46  & 0.64& 0.62  & 0.17 & 0.34 & 0.08\\
 & MSE & 0.23 & 0.43 & 0.40  & 0.04& 0.14 & 0.01\\
 \hline
\multirow{2}{*}{P=100}& Bias & 0.44 & 0.82& 0.78& 0.19 & 0.58 & 0.07\\
 & MSE & 0.21 & 0.69 & 0.63 & 0.05& 0.68 & 0.01\\
 \hline
\end{tabular}
\end{table}

\newpage

\subsection*{Appendix D}

\begin{figure}[h]
\centering
 \includegraphics[width=0.45\textwidth]{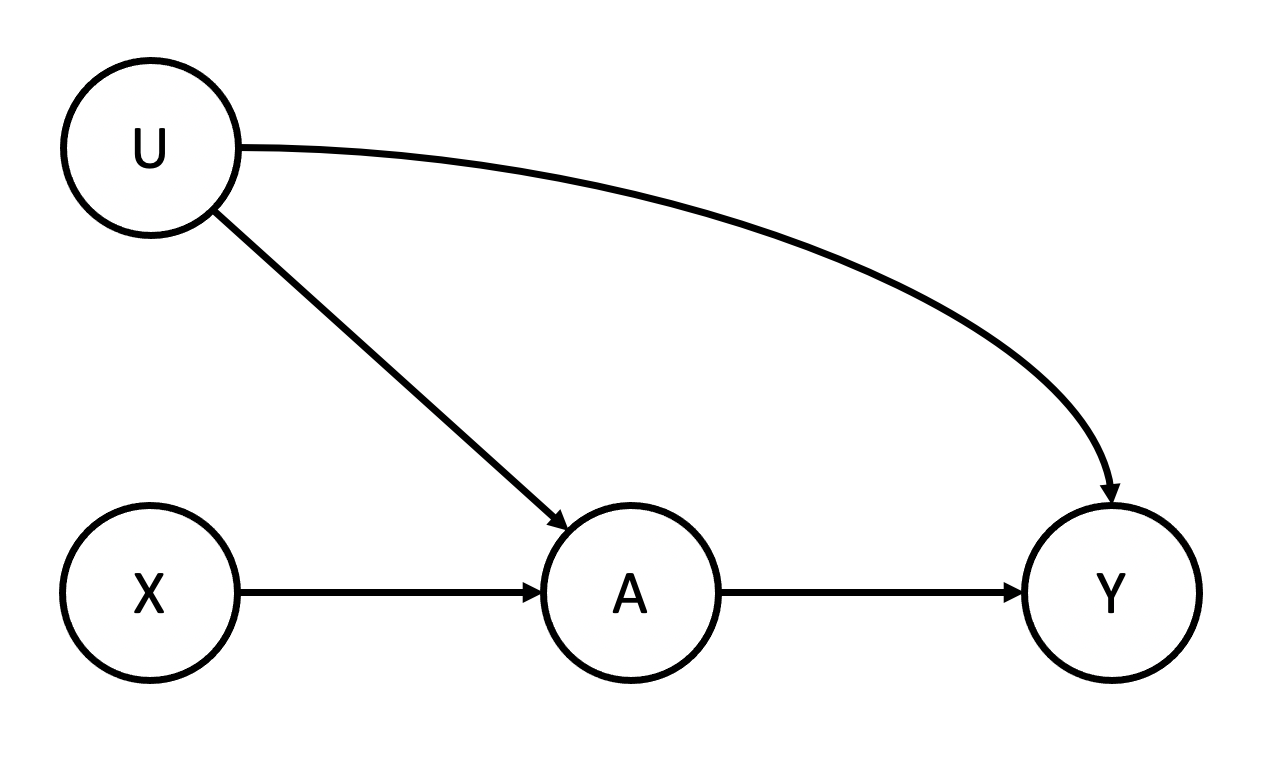}
    \caption{\footnotesize An illustration of ``Z-bias''. Variable $U$ is unmeasured and $X$ (instrument) affects outcome $Y$ only through $A$.}    \label{fig:zbias} 
\end{figure}

\begin{table}[h]
 \caption{\footnotesize The selected census and weather variables as confounders in each and both models with respect to posterior inclusion probability 100\%. Weather variables with varying displacements in the eight cardinal directions (N, NE, E, SE, S, SW, W, NW) are classified as distinct confounders. }
    \label{tab:pip}
    \centering
    \scriptsize
    \renewcommand\baselinestretch{1.7}\selectfont
    \begin{tabularx}{\textwidth}{|X||X|X|X|}
    \hline
\centering  \bf Variable Category  &\centering \bf In Marginal Model Only & \centering \bf In Both Models & {\centering \bf In Separate Model Only} \\
    \hline
        \hline
   \center Census &  {\bf (1 variable)} Gini Index  & {\bf(7 Variables)} Housing Units (urban), Pop. (Age $< 20$), Median Age, Pop (Bachelors \& Female), Median Income, Pop (Poverty Status) & {\bf(1 Variable)} Pop.(White)\\
    \hline
    \center   Local Weather  (Summer) & {\bf(1 Variable)} Planet Boundary Layer & {\bf(3 Variables)} Precipitation, Wind Speed, Temperature & {\bf(2 Variables)} Cloud Cover, Wind Angle \\
        \hline
    \center   Local Weather (Winter) & {\bf(-)} & {\bf(2 Variables)} Precipitation, Wind Angle & {\bf(4 Variables)} Humidity,  Cloud Cover, Wind Speed, Temperature  \\
    \hline
        \center   Regional Weather (Summer) & {\bf(5 Variables)} Planet Boundary Layer (E), Precipitation (NW, NE), Temperature (NE, E) & {\bf(13 Variables)} Humidity (NW, SE), Planet Boundary Layer (NW, SE, NE), Precipitation (W, N), Temperature (N, SE), Cloud Cover (SE, E) & {\bf(21 Variables)} Humidity (W, SW, S, NE, E), Cloud Cover (W, SW, NW, N, NE), Precipitation (SW, S, E), Temperature (W, SW, NW, S), Planet Boundary Layer (W, SW, S, N) \\
        \hline
      \center     Regional Weather (Winter) & {\bf(4 Variables)} Humidity (S, E), Cloud Cover (SW), Planet Boundary Layer (NW)   & {\bf(14 Variables)} Cloud Cover (SE, N), Humidity (SW, SE), Temperature (W, SW, SE, E), Precipitation (SW, NE, E), Planet Boundary Layer (N, NE, SE) & {\bf(18 Variables)} Cloud Cover (W, NW, S, NE, E), Precipitation (W, NW, S, SE), Humidity (W, NW, N), Planet Boundary Layer (W, SW, S, E), Temperature (NW, NE) \\
        \hline
       \centering Total Variables & \centering 11 & \centering 39 & {\centering \qquad \qquad \qquad 46}\\ 
    \hline
    \end{tabularx}
\end{table}

\begin{figure}[h]
\centering
 \includegraphics[width=0.75\textwidth]{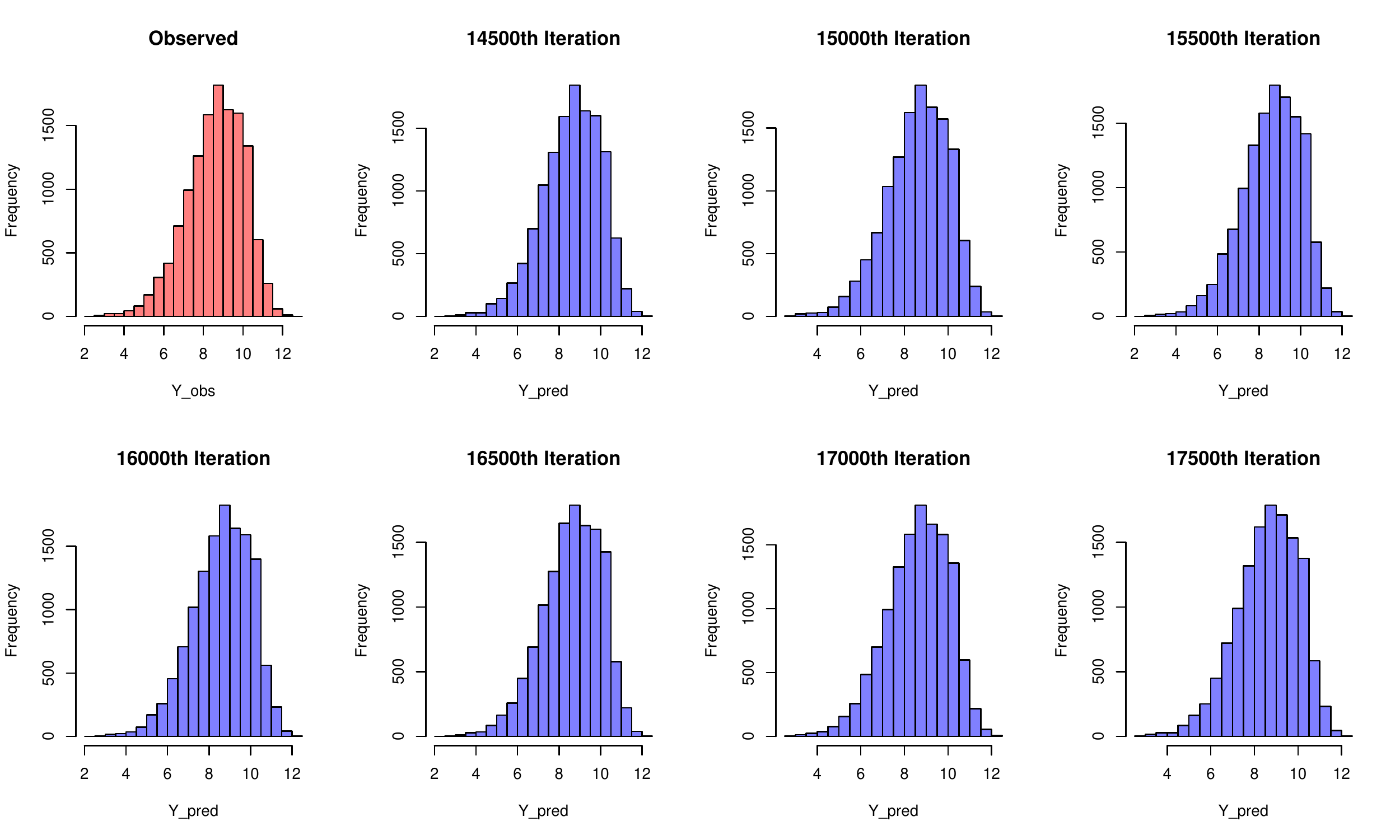}
 \includegraphics[width=0.75\textwidth]{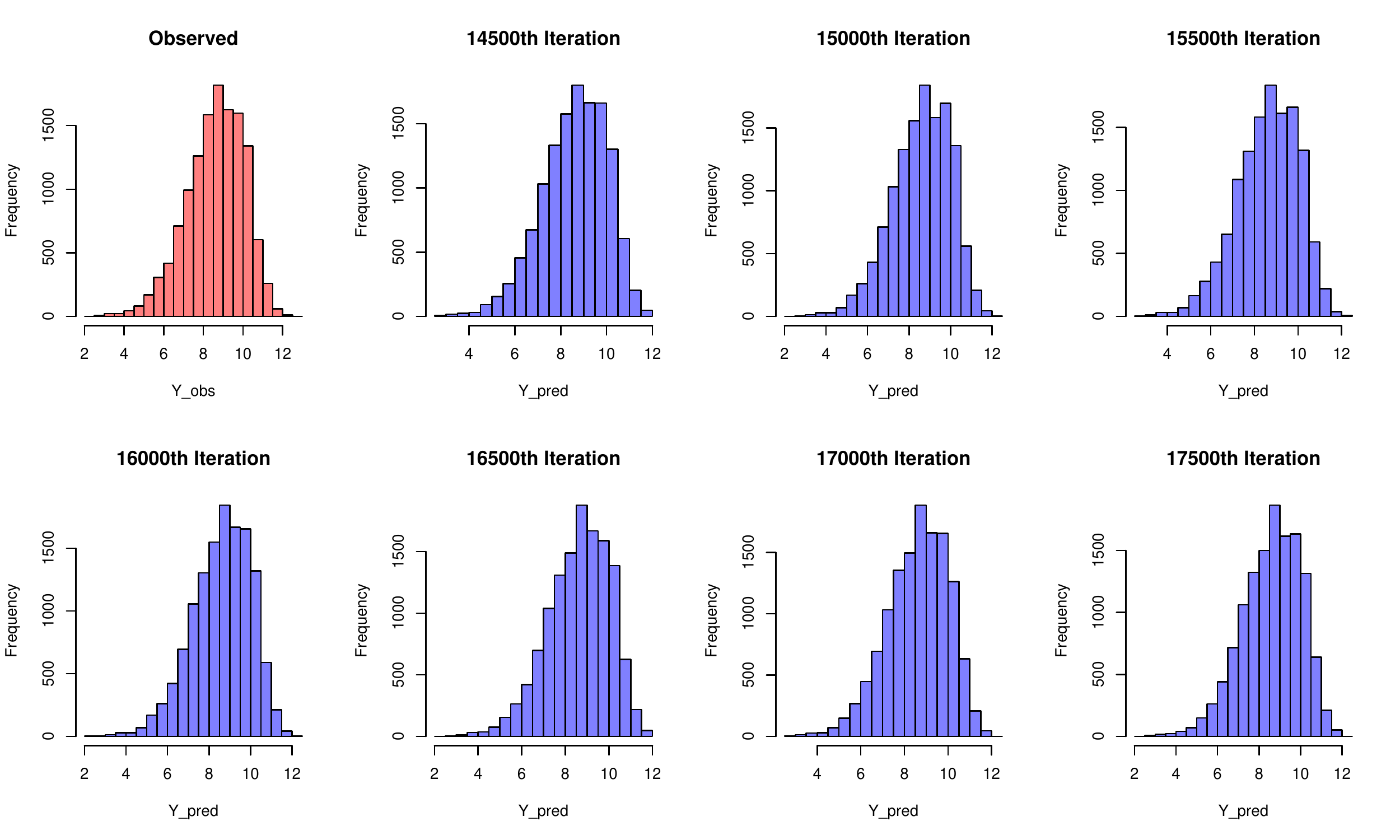}
    \caption{Posterior predictive checks. The first 2 rows: 7 replications of $Y$ from the posterior predictive distribution of the separate model. The last 2 rows: 7 replications of $Y$ from the posterior predictive distribution of the marginal model. }     \label{fig:posterior_predictive}
\end{figure}
\begin{figure}[t]
\centering
 \includegraphics[width=0.32\textwidth]{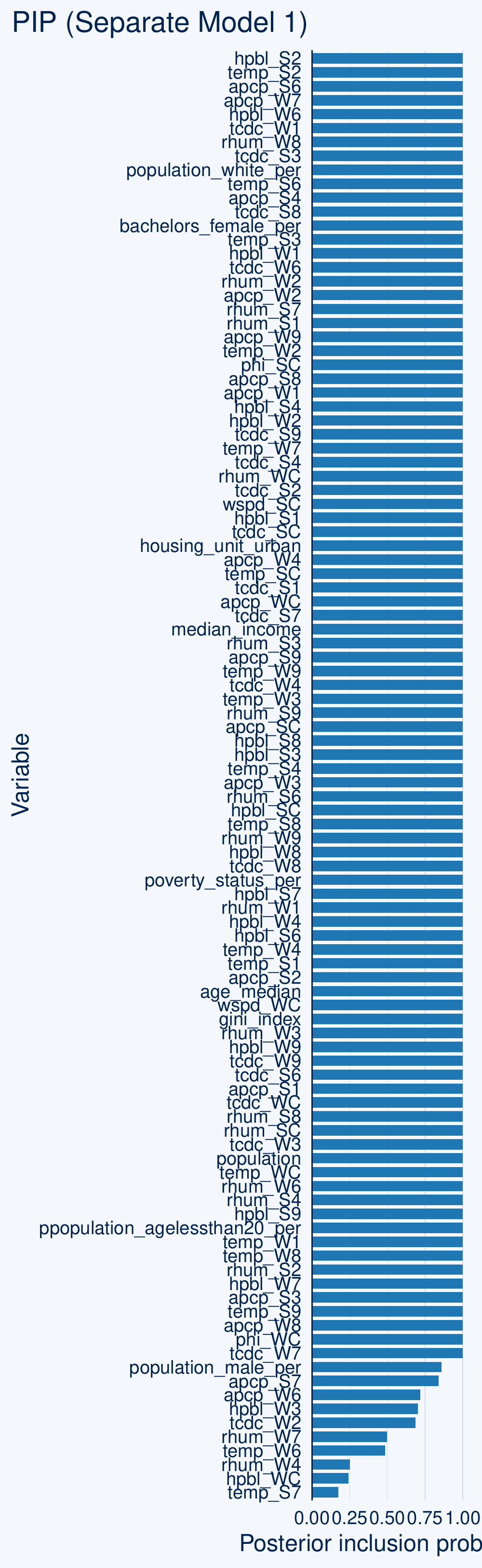}
 \includegraphics[width=0.32\textwidth]{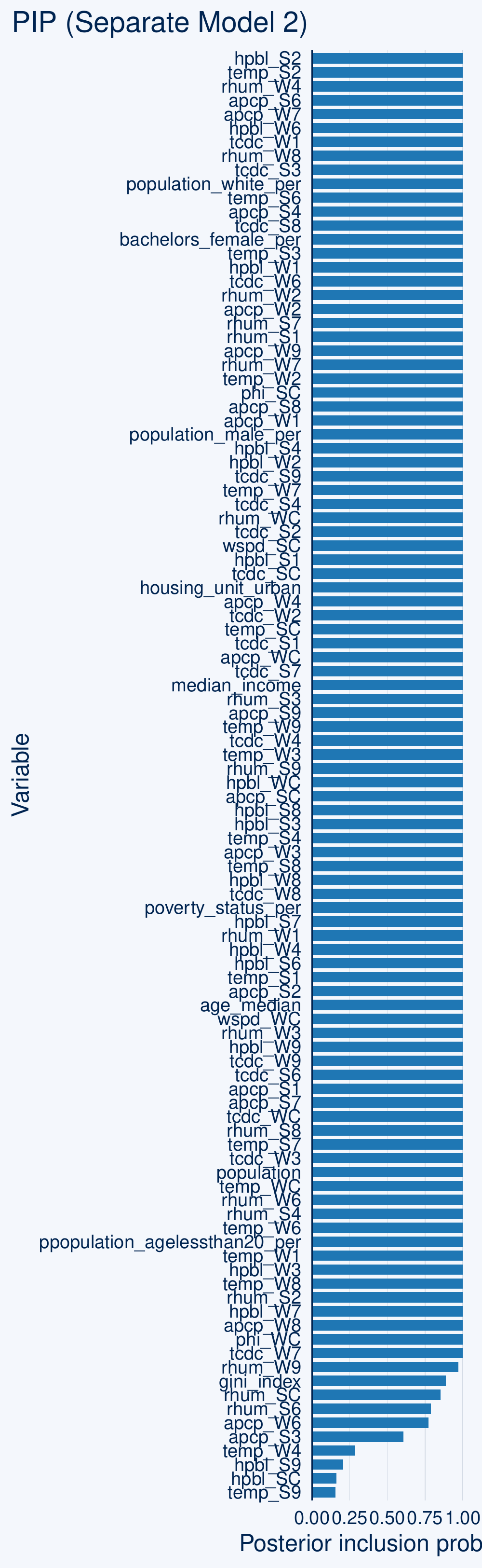}
     \includegraphics[width=0.32\textwidth]{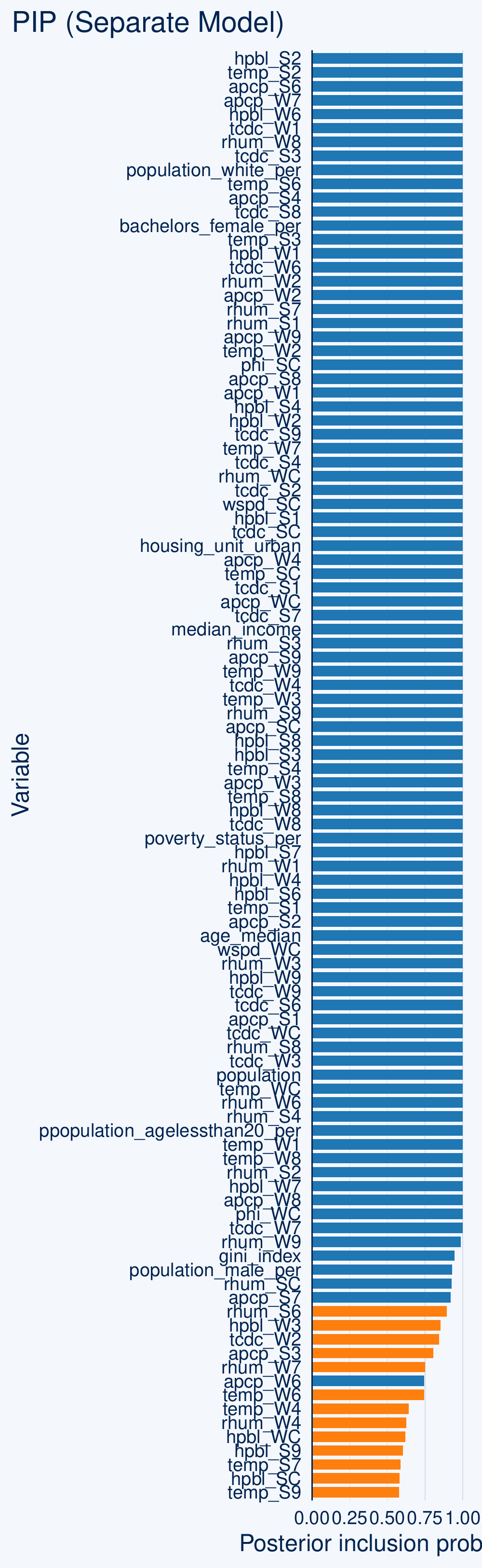}
\caption{The posterior probability of inclusion for the separate model.The first two plots are for two different MCMC chains and the last plot is for the average of two chains (orange bars indicate variables with different PIPs (difference larger than 20\%) in the two chains). The variable name of the form `$\text{name}_{\text{S1}}$' indicates the `name' variable in the area 100km to the west (1:West, 2:Southwest, 3: Northwest, 4: South, C:Center, 6:North, 7:Northeast, 8:Southeast, 9:East) in Summer (S:Summer; W:Winter). }     \label{fig:pip_sep}
\end{figure}

\begin{figure}[t]
\centering
 \includegraphics[width=0.32\textwidth]{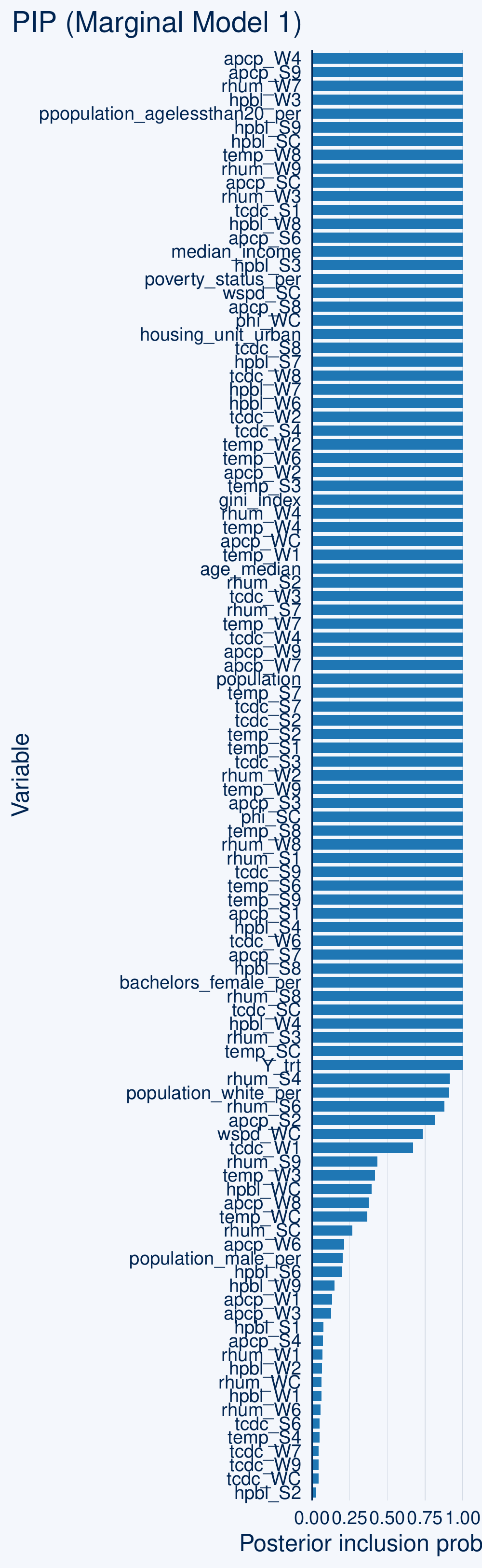}
 \includegraphics[width=0.32\textwidth]{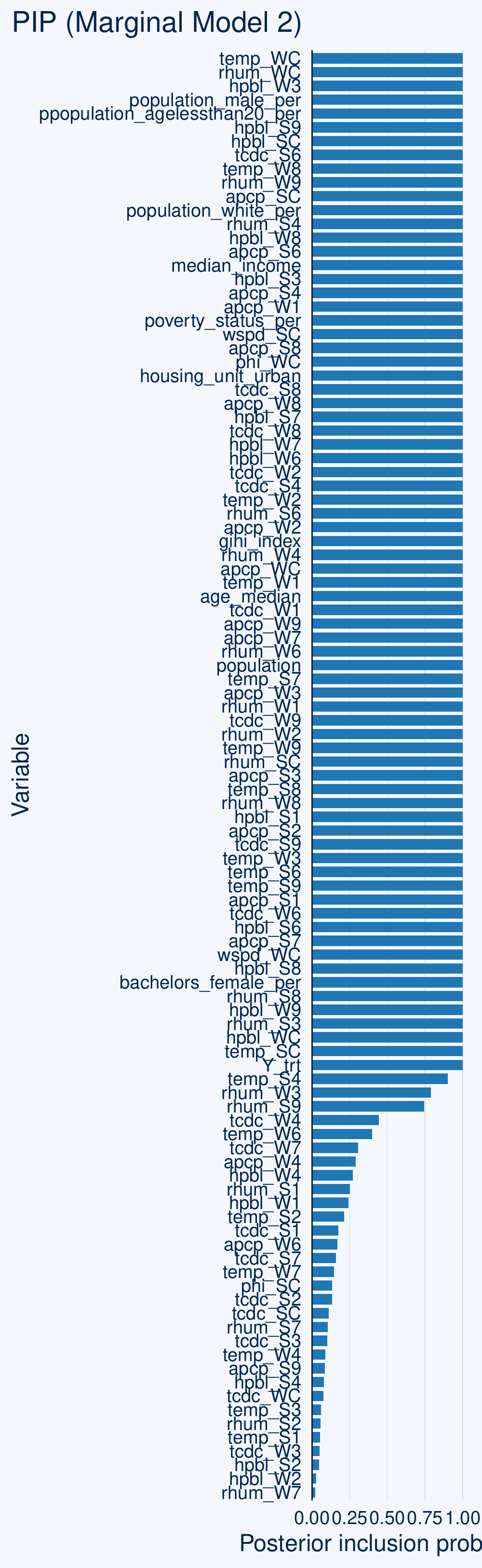}
     \includegraphics[width=0.32\textwidth]{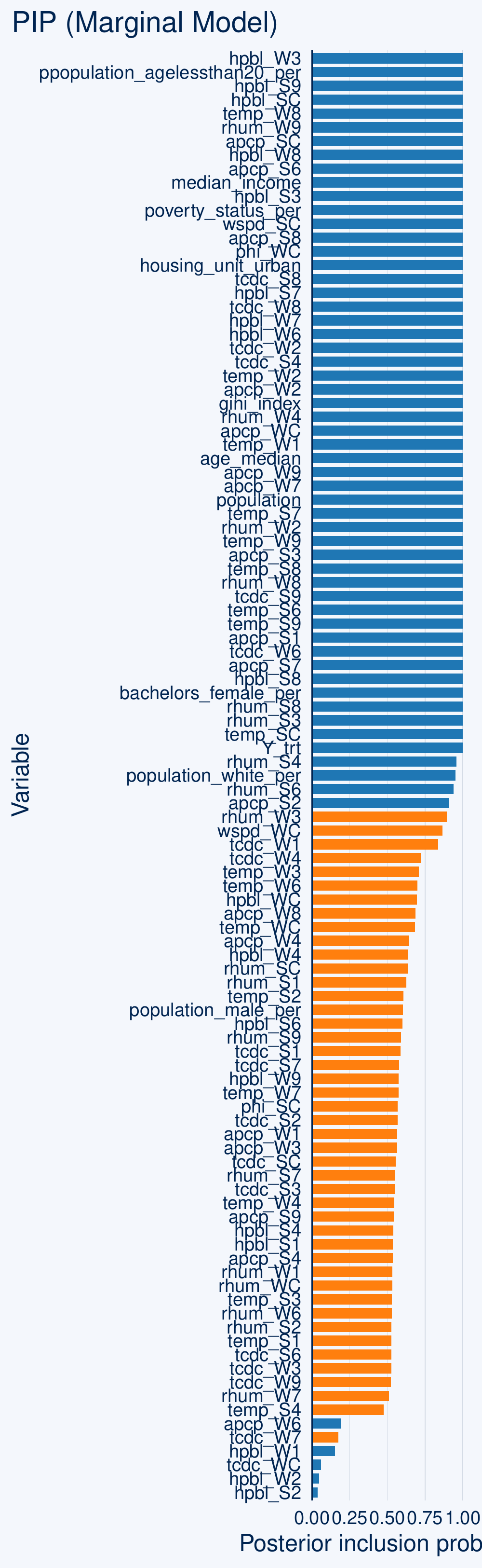}
\caption{The posterior probability of inclusion for the marginal model. The first two plots are for two different MCMC chains and the last plot is for the average of two chains (orange bars indicate variables with different PIPs (difference larger than 20\%) in the two chains). Note that the marginal model consider one additional variable (exposure) in the inclusion probabilities. The variable name of the form `NAME-S1' indicates the `name' variable in the area 100km to the west (1:West, 2:Southwest, 3: Northwest, 4: South, C:Center, 6:North, 7:Northeast, 8:Southeast, 9:East) in Summer (S:Summer; W:Winter). }     \label{fig:pip_mar}
\end{figure}

\begin{figure}[pt]
\centering
 \includegraphics[width=0.32\textwidth]{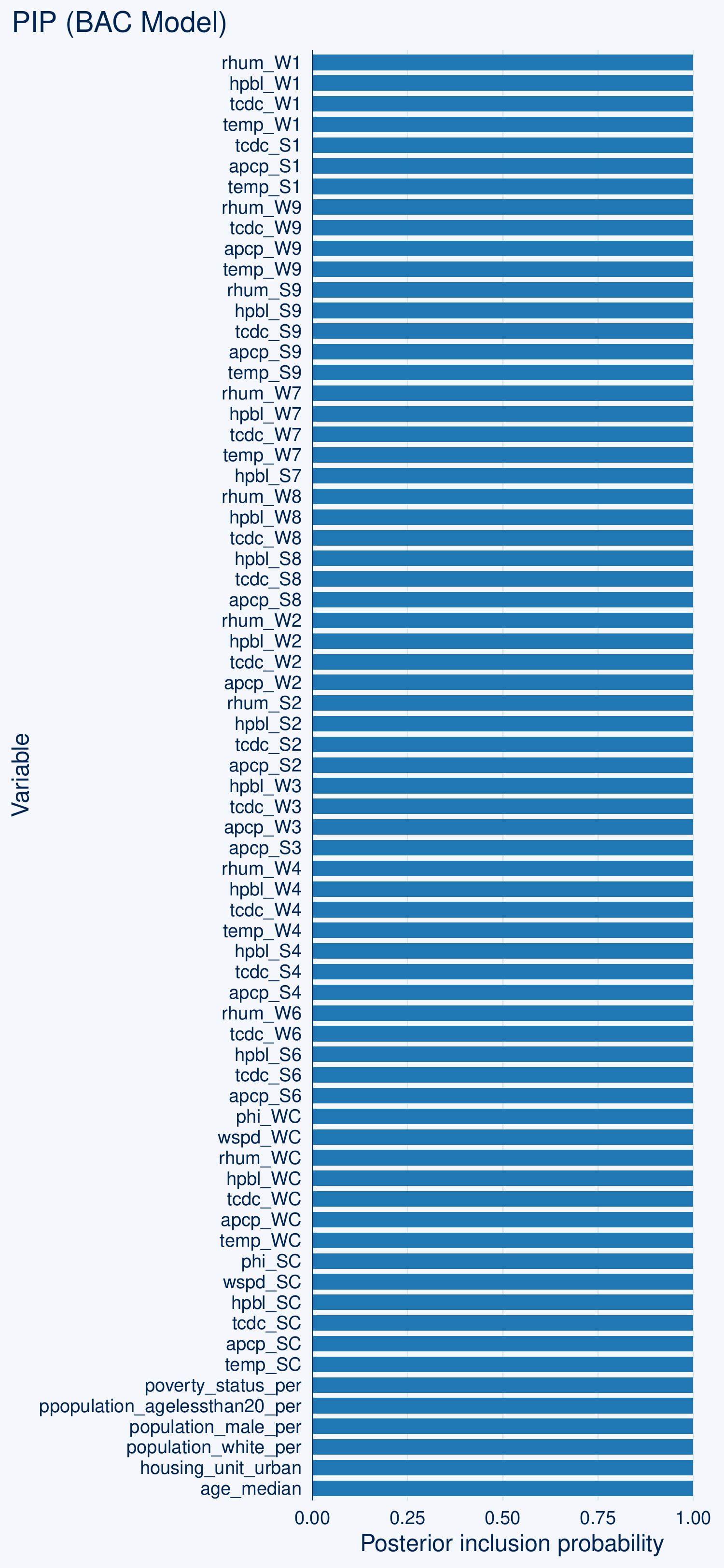}
  \includegraphics[width=0.32\textwidth]{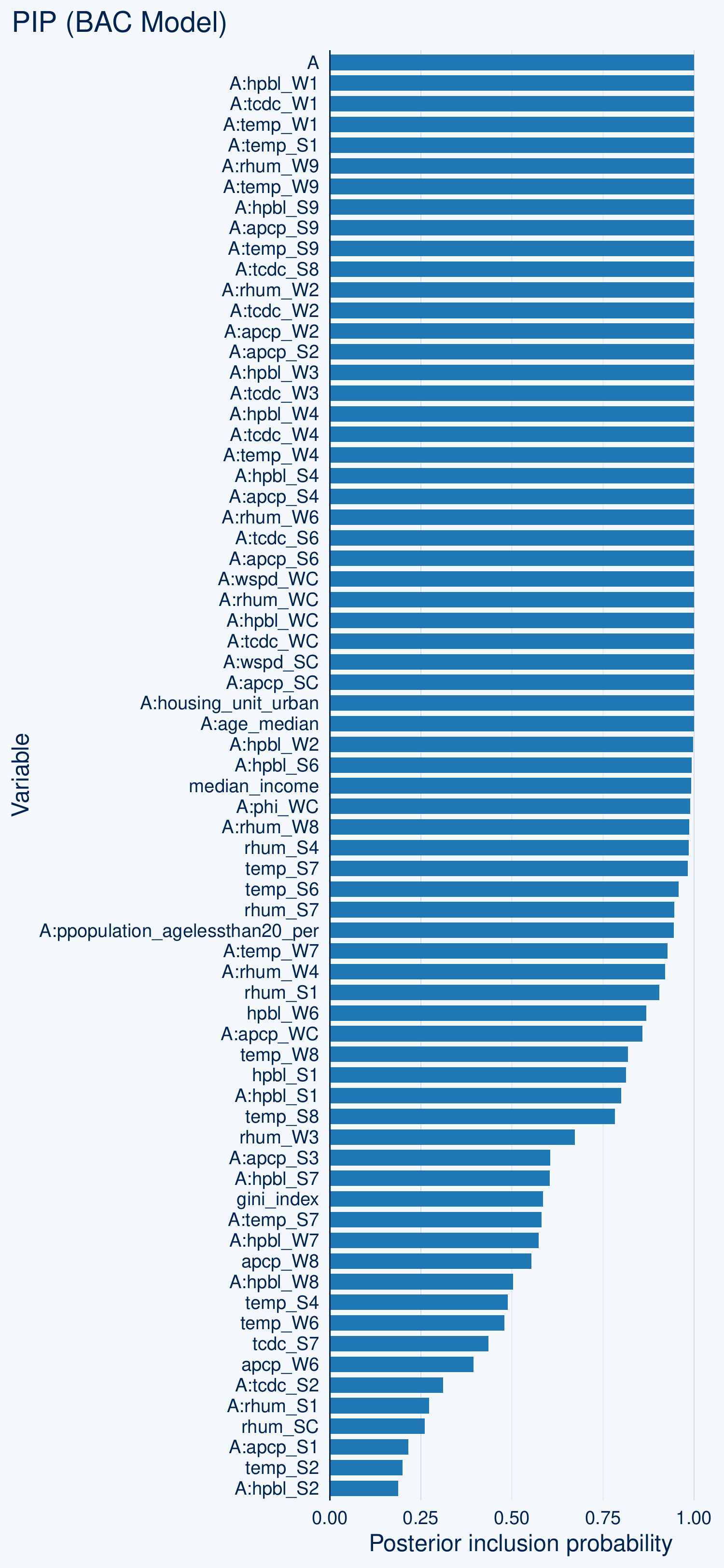}
 \includegraphics[width=0.32\textwidth]{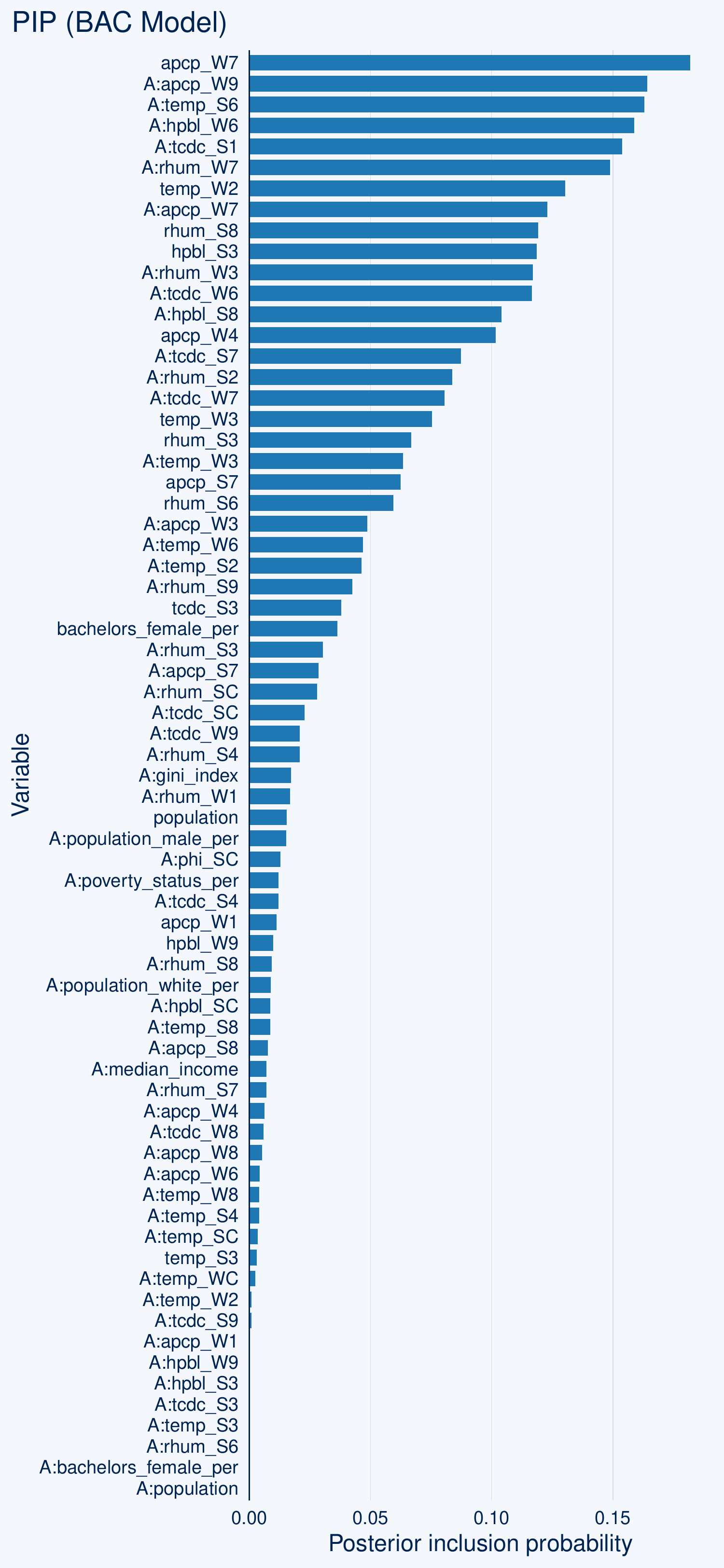}
    \caption{The posterior probability of inclusion for BAC from Wang et al. (2015). $A:X$ indicates an interaction term between exposure $A$ and variable $X$. The variable name of the form `$\text{name}_{\text{S1}}$' indicates the `name' variable in the area 100km to the west (1:West, 2:Southwest, 3: Northwest, 4: South, 5:Center, 6:North, 7:Northeast, 8:Southeast, 9:East) in Summer (S:Summer; W:Winter). }     \label{fig:bac}
\end{figure}

\begin{figure}[pt]
\centering
    \includegraphics[width=0.8\textwidth]{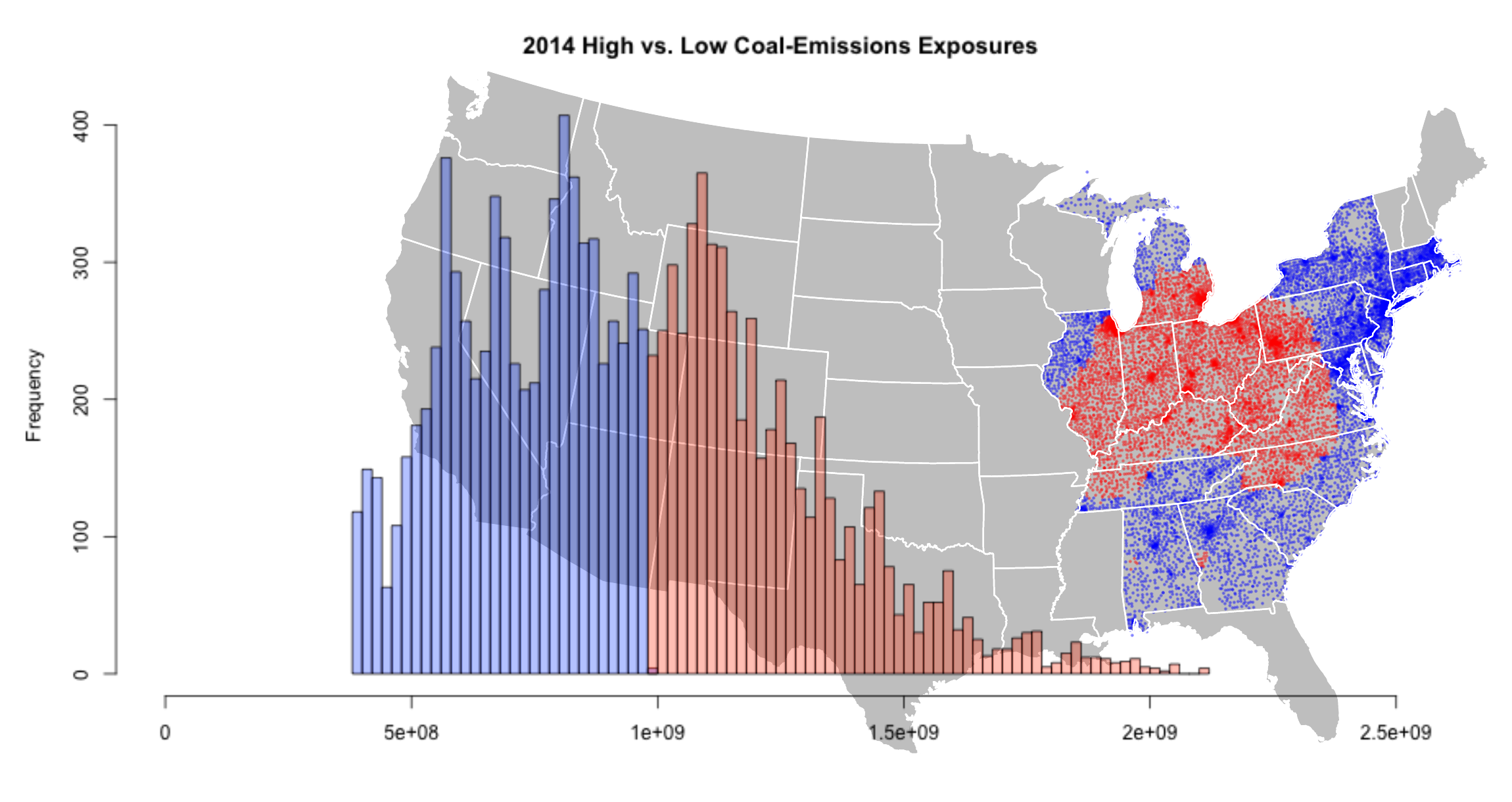}
\caption{The map of the zip code locations in 2014: the treated (low HyADS coal-emissions exposure; blue locations) versus the control (high HyADS coal-emissions exposure; red locations) locations. The analysis includes the zip codes located in the Eastern US. The histogram shows frequencies of high or low  HyADS coal-emissions exposures dichotomized at the mean value from the 2013 data (980249908). }\label{USmap2}
\end{figure}

\bibliographystyle{agsm}
\clearpage
\bibliography{BART_CS}